\documentclass[onecolumn,tightenlines,amsmath,amssymb,11pt,superscriptaddress,nofootinbib]{revtex4}

\usepackage{graphicx}
\usepackage{amsmath,amssymb,amsfonts,amsthm,stmaryrd,mathtools,bm,physics}
\usepackage{soul}
\usepackage{xcolor}
\usepackage{tikz}
\allowdisplaybreaks[1]
\usepackage[bookmarks,linktocpage, colorlinks=true, plainpages = false, citecolor = treegreen,  linkcolor=darkblue, urlcolor = darkblue, filecolor = blue]{hyperref} 

\definecolor{tealgreen}{rgb}{0.0, 0.5, 1.0}
\definecolor{darkblue}{rgb}{0., 0.4, 0.8}
\definecolor{cadmiumred}{rgb}{1., 0., 0.22}
\definecolor{treegreen}{rgb}{0., 0.7, 0.3}
\definecolor{emerald}{rgb}{0.31, 0.78, 0.47}
\definecolor{purple}{rgb}{1,0,1}
\definecolor{lime}{HTML}{A6CE39} 

\usepackage{pifont}

\def\be#1\ee{\begin{align}#1\end{align}}

\def\ba{\begin{eqnarray}}
\def\ea{\end{eqnarray}}

\usepackage[normalem]{ulem}

\begin{document}

\title{Violations of the null convergence condition in kinematical transitions between singular and regular black holes, horizonless compact objects, and bounces}

\author{Johanna Borissova}
\email{jborissova@pitp.ca}
\affiliation{Perimeter Institute for Theoretical Physics, 31 Caroline Street North, Waterloo, ON, N2L 2Y5, Canada}
\affiliation{Department of Physics and Astronomy, University of Waterloo, 200 University Avenue West, Waterloo, ON, N2L 3G1, Canada}
\author{Stefano Liberati}
\email{liberati@sissa.it}
\affiliation{SISSA, Via Bonomea 265, 34136 Trieste, Italy}
\affiliation{INFN, (Sezione di Trieste), Italy}
\affiliation{IPFU --- Institute for Fundamental Physics of the Universe, 
Via Beirut 2, 34014 Trieste, Italy}
\author{Matt Visser}
\email{matt.visser@sms.vuw.ac.nz}
\affiliation{School of Mathematics and Statistics, Victoria University of Wellington, PO Box 600, Wellington 6140, New Zealand}

\begin{abstract}

\bigskip

\noindent
{\sc Abstract:}\\ 
How do regular black holes evade the Penrose singularity theorem?
Various models of stationary regular black holes globally satisfy the null convergence condition (NCC). At first glance this might seem puzzling, as the NCC must generically be violated to avoid the focusing point implied by the Penrose theorem. 
{In fact, the Penrose singularity theorem depends on subtle global assumptions and {does not provide information about} where and when the singularity  actually forms. In particular inner horizons are typically reached at finite affine parameter, before null geodesic focusing occurs, and the region inside the inner horizon is not itself a trapped region. 
Specifically, the Bardeen, Dymnikova, Hayward
models} of stationary regular black holes feature an inner Cauchy horizon which violates global hyperbolicity, hence violating one of the key assumptions of Penrose's singularity theorem, and furthermore challenging their viability as long-living end-points of gravitational collapse.
In contrast, during non-stationary processes describing kinematic transitions between standard singular black holes and regular black holes or horizonless compact objects, the inner horizon --- when present --- need not act as a Cauchy horizon. 
This raises the intriguing possibility that the NCC might instead be violated during intermediate stages of such transitions. 
Our detailed analysis confirms that NCC violations occur frequently during such kinematic transitions, even when the stationary end-point spacetimes respect the NCC.
We also investigate analogous transitions toward black-bounce spacetimes and their horizonless compact counterparts, wormholes, where the NCC is always violated. 
These findings offer new insights into how regular black holes and related objects evade the constraints imposed by the  Penrose singularity theorem.

\bigskip
\LaTeX-ed \today.


\end{abstract}

\maketitle

\clearpage
\tableofcontents

\clearpage
\section{Introduction}

How do regular black holes (RBHs) evade the Hawking--Penrose singularity theorems~\cite{Penrose:1964wq, Hawking:1970zqf} (see also~\cite{Senovilla:2014gza,Hawking:1973uf})? This question is more subtle than one might suppose. In this paper we will focus on the Penrose (1965) singularity theorem~\cite{Penrose:1964wq} and treat the Hawking--Penrose (1970) singularity theorem~\cite{Hawking:1970zqf} in a companion paper~\cite{ToAppear}. The Penrose theorem establishes the null geodesic incompleteness of a spacetime $(\mathcal{M},g)$ under the following conditions:
\begin{itemize}
\item[i)]{null convergence condition (NCC): $R_{\mu\nu}\,\ell^\mu \ell^\nu \geq 0 $ for all null vectors $\ell^\mu$,}
\item[ii)]{there is a non-compact Cauchy hypersurface $\Sigma$ in $\mathcal{M}$,}
\item[iii)]{there is a closed trapped surface $\mathcal{T}$ in $\mathcal{M}$\,.}
\end{itemize}
It is easy to check that the most common regular black holes (Bardeen, Dymnikova, Hayward, and close variants thereof~\cite{Bardeen:1968bh, Dymnikova:1992ux, Hayward:2005gi}) globally satisfy the NCC.\footnote{More distantly related regular black holes, such as the Simpson--Visser black bounce~\cite{Simpson:2018tsi} (or the closely related Simpson--Visser naked traversable wormhole) do, however, globally violate the NCC. 
In contrast, the Minkowski-core regular black hole~\cite{Simpson:2019mud,Berry:2020ntz,Simpson:2021dyo,Simpson:2021zfl} exhibits intermediate behaviour --- the NCC is satisfied deep in the core and violated in the periphery.}
We focus on the geometric NCC, rather than the implicitly dynamical null energy condition (NEC) since the NEC makes dynamical assumptions regarding the applicability of the Einstein equations --- we shall instead phrase the discussion in purely geometrical terms, eschewing the need for any particular choice of dynamics~\cite{Carballo-Rubio:2019fnb}. 
(For general discussions of energy conditions, in contrast to what can be done at a purely kinematic level, see, for instance,~\cite{Curiel:2014zba, Borde:1987qr,Barcelo:2002bv,Visser:1999de,Martin-Moruno:2017exc, Martin-Moruno:2013wfa,Martin-Moruno:2013sfa,Martin-Moruno:2015ena}.) 
Since the most commonly occurring regular black holes satisfy the NCC, they must evade the Penrose singularity theorem in some other manner --- the only remaining point of evasion being the hypothesized {(global)} causality condition which postulates the existence of a non-compact Cauchy hypersurface, tantamount to demanding global hyperbolicity Specifically, the presence of a static or stationary inner horizon, which is thereby a Cauchy horizon, violates global hyperbolicity. 

We emphasize that the Hawking--Penrose (1970) singularity theorem~\cite{Hawking:1970zqf} (see also~\cite{Senovilla:2014gza,Hawking:1973uf}) explicity removes the assumption of existence of a Cauchy hypersurface by instead demanding the generic condition for the curvature tensor and the absence of closed timelike curves, but most importantly assumes the timelike convergence condition (TCC) {\it in addition} to the NCC. Concerning the Hawking--Penrose theorem (1970), it is this last condition which is violated by the standard de Sitter core regular black holes.

The above observations apply also to the much more familiar and singular Reissner--Nordstr\"om spacetime. It is perhaps a little sobering to realize that, due to the presence of a Cauchy horizon, the standard Penrose singularity theorem does not apply. The Hawking--Penrose theorem, however, does establish the geodesic incompleteness of the Reissner--Nordstr\"om spacetime. Its regular counterpart, the Bardeen black hole~\cite{Bardeen:1968bh}, evades this theorem purely by the fact that it violates the TCC. We analyze such TCC violations in a companion paper~\cite{ToAppear}. 

Regarding the Penrose (1965) theorem, in non-stationary spacetimes the situation is much more complicated and interesting.
If, for example, one constructs a kinematic model of a transition from/to a regular black hole to/from either a standard (singular) black hole, or to/from some horizonless compact object (HCO), then the evolving inner horizon (when present) need not be a Cauchy horizon. Similarly, in a dynamical gravitational collapse leading to the creation of a Kerr black hole only an inner horizon will {need} to be considered at {finite} times, at least before the formation of a time-like singularity, (which will {then also} imply the formation of a Cauchy horizon).

If the inner horizon is not Cauchy, then there is no {\em a priori} reason to not apply the Penrose singularity theorem and hence the avoidance of a singularity implied by a late-time realization of one of the stationary regular black-hole geometries naturally leads to the conclusion that the NCC most probably has to be violated {\em during the kinematic transition}, though not at the stationary end-point.
In this work, we shall verify that in several scenarios this is actually what happens: NCC violation can and does typically occur during the kinematic transition, even between end-point spacetimes that do themselves satisfy the NCC.

A comment is in order to better justify our {specific} choice of {examples}. Regular black holes are typically assumed to arise through some form of quantum-gravity-induced regularization that prevents the formation of a focusing point for the outgoing null congruence.\footnote{Remember that in the trapped region ``outgoing'' means ``less ingoing''.} This avoidance of focusing points is a crucial step in the framework of the Hawking–Penrose singularity theorems and plays a key role in predicting the classes of regular solutions that replace the traditional singular black holes~\cite{Carballo-Rubio:2019fnb,Carballo-Rubio:2019nel,Carballo-Rubio:2018pmi}. Consequently, the most relevant transition for our study would involve a stellar-like object collapsing into one of these regular solutions.

Unfortunately, such transitions require detailed knowledge of the dynamics governing the process --- knowledge that is currently unavailable. For this reason, in this work, we do not consider fully realistic collapse scenarios. Instead, we model transitions corresponding to the regularization of pre-existing spherically symmetric, static, singular solutions, such as the Schwarzschild black hole, or metrics that, while still featuring a curvature singularity, are at least continuous at $r=0$, i.e.~black holes with integrable singularities, which allows us to verify that any potential artefacts, introduced by an initially ill-defined metric at the origin, do not affect the nature of the transition.

Additionally, we explore the only type of transition between a regular black hole and a horizonless object that can be easily modeled: the transition to their horizonless compact counterparts~\cite{Carballo-Rubio:2022nuj}. This provides a framework for analyzing how regular black holes might connect to alternative compact configurations and sheds light on the underlying mechanisms.

Although our analysis is purely classical, our results have important implications for the quantum theory associated with the formation of regular black holes and bouncing spacetimes. The necessary NCC violation observed during the transition from a singular state to a regular state can be expected, in any physically realistic scenario, to have an intrinsically quantum origin. While computing the renormalized stess-energy tensor (RSET) in such dynamical situations remains a challenging and unsolved task, our analysis reveals necessary properties required of the RSET in order to induce the formation of a regular final state.

This article is structured as follows. In Section~\ref{Sec:NCCGeneral} we explicitly derive the conditions imposed by the NCC in a general dynamical spherically symmetric spacetime. Furthermore  we distinguish between the two physically most relevant choices of radial functions suitable for describing either black holes in Subsection~\ref{SecSub:Rr}, or bouncing spacetimes in Subsection~\ref{SecSub:Rr0}. In Section~\ref{Sec:KinematicalTransitionModels} we analyze violations of the NCC in kinematical transition models between two stationary geometries. To that end, in Subsection~\ref{SecSub:InterpolatingFunctions} we first introduce smooth interpolating functions to model such transitions. Subsequently, we analyze violations of the NCC in transitions from singular to regular black holes in Subsection~\ref{Sec:SBHtoRBH}, from regular black holes to horizonless compact objects in Subsection~\ref{Sec:RBHtoUCO}, from singular black holes to bouncing or wormhole spacetimes in Subsection~\ref{Sec:SBHtoBounce}, and from bouncing to wormhole spacetimes in Subsection~\ref{Sec:SVBHtoSVWormhole}.  We finish with a discussion in Section~\ref{Sec:Discussion}.

\section{Null convergence condition in spherically symmetric spacetimes}\label{Sec:NCCGeneral}

Consider a spherically symmetric spacetime with line element in ingoing Eddington--Finkelstein 
coordinates
\be\label{eq:MetricImplodingSphericalSymmetry}
\dd{s^2} = -f(r,v)\dd{v}^2 + 2  \dd{v}\dd{r} + R(r,v)^2   \dd{\Omega}^2 \quad \text{with} \quad f(r,v) = 1- \frac{2m(r,v)}{R(r,v)}\ .
\ee
Here $\dd{\Omega}^2$ is the area element and we assume the function $R(r,v)$ to be non-negative. 
The null convergence condition (NCC) requires that
\be
R_{\mu\nu}\;\ell^\mu \ell^\nu \geq 0 \quad\text{for {\it any} null vector} \,\,\,\ell^\mu.
\ee
Let us first introduce the two radial null vectors
\be\label{eq:kl}
k^\mu = - \partial_r = (0,-1,0,0) \,\,\,\quad \text{and} \,\,\,\quad l^\mu = \partial_v + \frac{f}{2}\partial_r =  \qty(1,\frac{f}{2},0,0)\,\,\, 
\ee
tangent to ingoing and ``outgoing'' radial null geodesics respectively.\footnote{In the trapped region, $f<0$, {one has $l^r <0 $, and } the ``outgoing'' radial null geodesics are merely the ``less ingoing'' of the two radial null geodesics. }
It is then easy to check that $g(k,k)=0=g(l,l)$ and $g(k,l)=-1$.\footnote{With these conventions the ingoing null geodesics are affine parameterized, $k^\nu \; \nabla_\nu k^\mu =0$, whereas the outgoing null geodesics are not affine parameterized, $l^\nu \; \nabla_\nu l^\mu = {1\over2} f'(r,v)\;l^\mu \neq 0$.}
In the discussion below we will primarily investigate the NCC inequalities for the two radial null vectors $k^\mu$ and $l^\mu$ defined above.  Contracting these radial null vectors with the Ricci tensor leads to 
\ba
R_{\mu\nu}\; k^\mu k^\nu &=& - \frac{2 R''}{R}\,,\label{eq:NCCk}\\
R_{\mu\nu}\; l^\mu l^\nu &=&- \frac{1}{2  R}\qty(  2\dot{f}R' + f^2 R'' + 4 f \dot{R}' - 2f'\dot{R} +4  \ddot{R}  )\,, \label{eq:NCCl}
\ea
where primes and dots denote partial derivatives with respect to $r$ and $v$. Note that the first expression corresponding to an ingoing radial null geodesic does not involve any time derivatives. The second expression, corresponding to ``outgoing'' radial null geodesics, is more subtle.

For completeness note that in spherical symmetry any arbitrary vector can without loss of generality be expressed in the form $\ell^\mu =(\alpha,\beta,\gamma,0)$, whence imposing the null condition and rescaling $\alpha\to1$ implies
\ba
\ell^\mu = \left(1, {f(r,v)- \gamma^2R(r,v)^2\over 2}, \gamma,0\right) 
\ea
with $g(\ell,\ell)=0$;  $g(\ell,k) = -1$; and $g(\ell,l) = - {R(r,v)^2 \gamma^2\over 2}$.

Note that $\lim_{\gamma\to0} \ell^\mu = l^\mu$ and $\lim_{\gamma\to\infty } [\gamma^{-2} \ell^\mu]  =(0, -{1\over2} R^2,0,0) \propto k^\mu$, so this generic null vector smoothly interpolates between the two radial null vectors.
The general quantity of interest in the NCC, $R_{\mu\nu}\; \ell^\mu \ell^\nu$, is then easily computed to be:
\ba
R_{\mu\nu}\; \ell^\mu \ell^\nu =  [R_{\mu\nu}\; l^\mu l^\nu]
+{\gamma^4 R(r,v)^4\over 4} [R_{\mu\nu}\; k^\mu k^\nu]
+{\gamma^2\over2}\left( R^2 f'' -2 f\; [R']^2- 4 R' \dot R +2 \right).
\ea
Therefore, even in the most general null directions, major insights originate from evaluating the NCC in the two radial null directions.
(In the discussion below we will primarily be interested in establishing violations of the NCC, so it will often be sufficient to restrict attention to the specific quantity $R_{\mu\nu}\; l^\mu l^\nu$ corresponding to ``outgoing'' radial null geodesics.)
Now we will distinguish two classes of geometries.

\subsection{Spacetimes with $R(r,v) = r$}
\label{SecSub:Rr}

For $R(r,v) = r$ the expression in~\eqref{eq:NCCk} vanishes, $R_{\mu\nu}\; k^\mu k^\nu\to 0$, whereas~\eqref{eq:NCCl} simplifies to
\ba
R_{\mu\nu}\; l^\mu l^\nu = - \frac{\dot{f}}{ r} = \frac{2\dot{m}}{r^2}\,.\label{eq:NCClRr}
\ea
We see that in any (instantaneously) static spacetime ($\dot{m}=0$) the NCC associated with the radial null vectors $k^\mu$ and $l^\mu$ is always marginally satisfied. This holds for {\it any}\, static spherically symmetric black hole for which $R(r,v)=r$, i.e., in particular for the singular Schwarzschild black hole equally well as for regular black holes with smooth de Sitter cores, such as the Dymnikova~\cite{Dymnikova:1992ux}, Bardeen~\cite{Bardeen:1968bh} or Hayward~\cite{Hayward:2005gi} black holes. This observation motivates us to investigate kinematical models for the formation of these geometries during a transition from a singular black hole and to analyze possible necessary violations of the NCC. In any non-static situation it follows from~\eqref{eq:NCClRr} that the (radial) NCC will be violated if and only if $\dot{m}<0$.

Even in a completely general null direction $\ell^\mu = \left(1, {1\over2}[f-\gamma^2 r^2],\gamma,0\right)$ we see
\ba
R_{\mu\nu}\; \ell^\mu \ell^\nu = -  \frac{\dot{f}}{ r} 
+ \gamma^2 \left(1-f +{r^2 f''\over2}\right)=  \frac{2\dot{m}}{r^2} + \gamma^2\left(2 m' - r m''\right)\,.\label{eq:NCClRr-generic}
\ea
The NCC requires that for all $\gamma$ we have $R_{\mu\nu}\; \ell^\mu \ell^\nu\geq 0$, whence for $R(r,v)=r$ we deduce
\ba
\mathrm{NCC}: \quad \dot m\geq 0; 
\qquad \mathrm{and} \qquad    
\left(2 m' - r m''\right) \geq 0\,.
\ea
If (without prejudice as to any specific choice of dynamics) one chooses to define a function $\rho(r,v)$ by $m'(r,v) = 4\pi \rho(r,v) r^2$, then one sees 
$\left(2 m' - r m''\right) = -4\pi \rho'(r,v)\,r^3 $ and so 
in the current context the  NCC in a generic null direction can be more compactly written as 
\ba
\mathrm{NCC}: \quad \dot m\geq 0; 
\qquad \mathrm{and} \qquad    
\rho' \leq 0\,.
\ea

\subsection{Spacetimes with $R(r,v)$ having non-zero local minimum at $r=r_0$}\label{SecSub:Rr0}

 If $\eval{R(r,v)}_{v = \text{const}}$ as a function of $r$ has a strict non-zero local minimum at some $r=r_0$, then the assumption that $R$ is non-negative implies $R_0\equiv R({r_0}) >0$, in addition to both $R_0' =0$ and $R_0''>0$. (Geometrically, this corresponds to a wormhole throat.) Consequently the expression in~\eqref{eq:NCCk} evaluated at $r_0$ simplifies to
\ba
\eval{R_{\mu\nu}\; k^\mu k^\nu}_{r=r_0} = - \frac{2 R_{0}''}{ R_0}\label{eq:NCCkRr0}\,.
\ea
Independent of the kinematics, the NCC associated with the null vector $k^\mu$ tangent to ingoing null geodesics is always violated at least locally near $r=r_0$.\footnote{More rarely, one might instead wish to consider $R(r,v)$ having a non-zero local maximum at $r=r_0$. Geometrically this would correspond to an anti-throat, a local maximum in the cross-sectional area of a ``bag of gold'' spacetime~\cite{Fu:2019oyc, Marolf:2008tx,
N-O-Murchadha_1987}. The NCC would be satisfied at such a local maximum.} In a static spacetime where moreover $\dot{R}=0$ holds, the NCC associated with the null vector $l^\mu$ tangent to outgoing null geodesics is also violated locally near $r=r_0$, as can be seen from equation~\eqref{eq:NCCl}:
$R_{\mu\nu}\; l^\mu l^\nu \to  - f^2 R''_0/(2 R_0)$. These observations apply in particular to static bouncing or wormhole geometries. A wormhole throat is defined such that the cross-sectional area of a bundle of light rays passing orthogonally through it has a strict local non-zero minimum. Therefore, if the throat is located at $r=r_0$, the function $R$ will satisfy the assumptions of this subsection. Static geometries with these properties in this sense by definition violate the NCC. This leads to a defocusing of both ingoing and outgoing null geodesics and ensures that the assumptions of the Penrose singularity theorem~\cite{Penrose:1964wq} are violated. Thereby bouncing or wormhole geometries can be free from spacetime singularities. We will explore the NCC violation expressed by equation~\eqref{eq:NCCkRr0} in the context of kinematical models which interpolate between a black hole with $R(r,v)=r$ and a wormhole or bouncing geometry.

Still in a static spacetime, in a completely general null direction, $\ell = \left(1, {1\over2}[f-\gamma^2 R^2],\gamma,0\right)$, at the throat one has
\ba
R_{\mu\nu}\; \ell^\mu \ell^\nu \to   - \left( f_0^2 
+{\gamma^4 R_0^4} \right) \frac{R_{0}''}{ 2 R_0}
+{\gamma^2\over2}\left( R_0^2 f_0''  +2 \right).
\ea
For both $\gamma\to0$ and $\gamma\to\infty$ (the two radial null directions) we see that it is the sign of $R_0''$ that controls whether or not the NCC is violated at the throat/anti-throat.

\section{Kinematical transition models}\label{Sec:KinematicalTransitionModels}

 In this section we will investigate possible violations of the NCC in kinematical processes describing transitions from a singular to a regular black hole or from a regular black hole to a horizonless compact object satisfying the assumptions of Section~\ref{SecSub:Rr}. Subsequently we will demonstrate the NCC violation in transitions from a singular black hole satisfying the assumptions of Section~\ref{SecSub:Rr} to a regular bouncing or wormhole geometry satisfying the assumptions of Section~\ref{SecSub:Rr0}. To that end we first need to specify how we interpolate in time between two static geometries. 

\subsection{Interpolating mass and radial functions}\label{SecSub:InterpolatingFunctions}

\begin{figure}[!h]
	\centering
	\includegraphics[width=.48\textwidth]{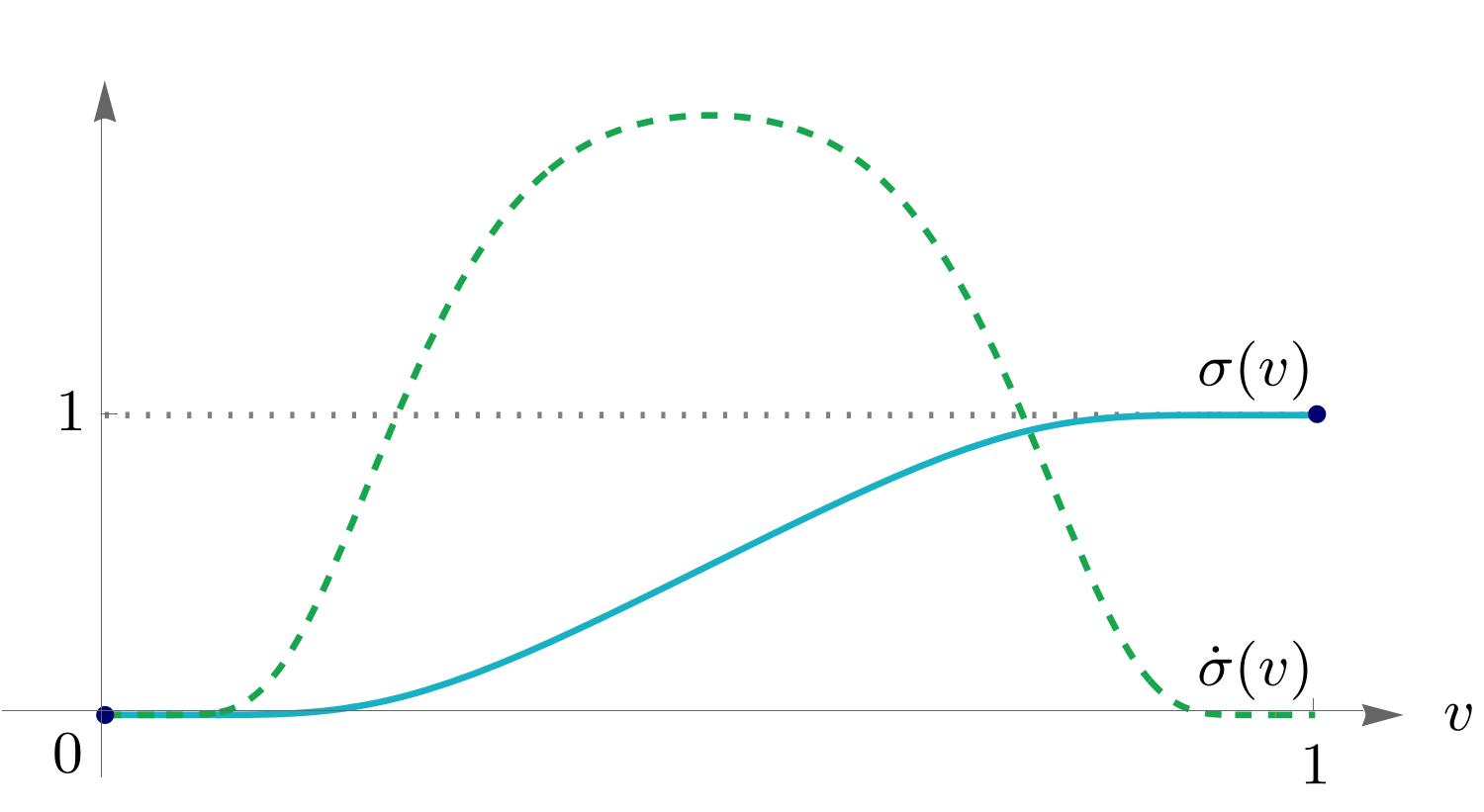}
	\hspace{0.4cm}
	\includegraphics[width=.48\textwidth]{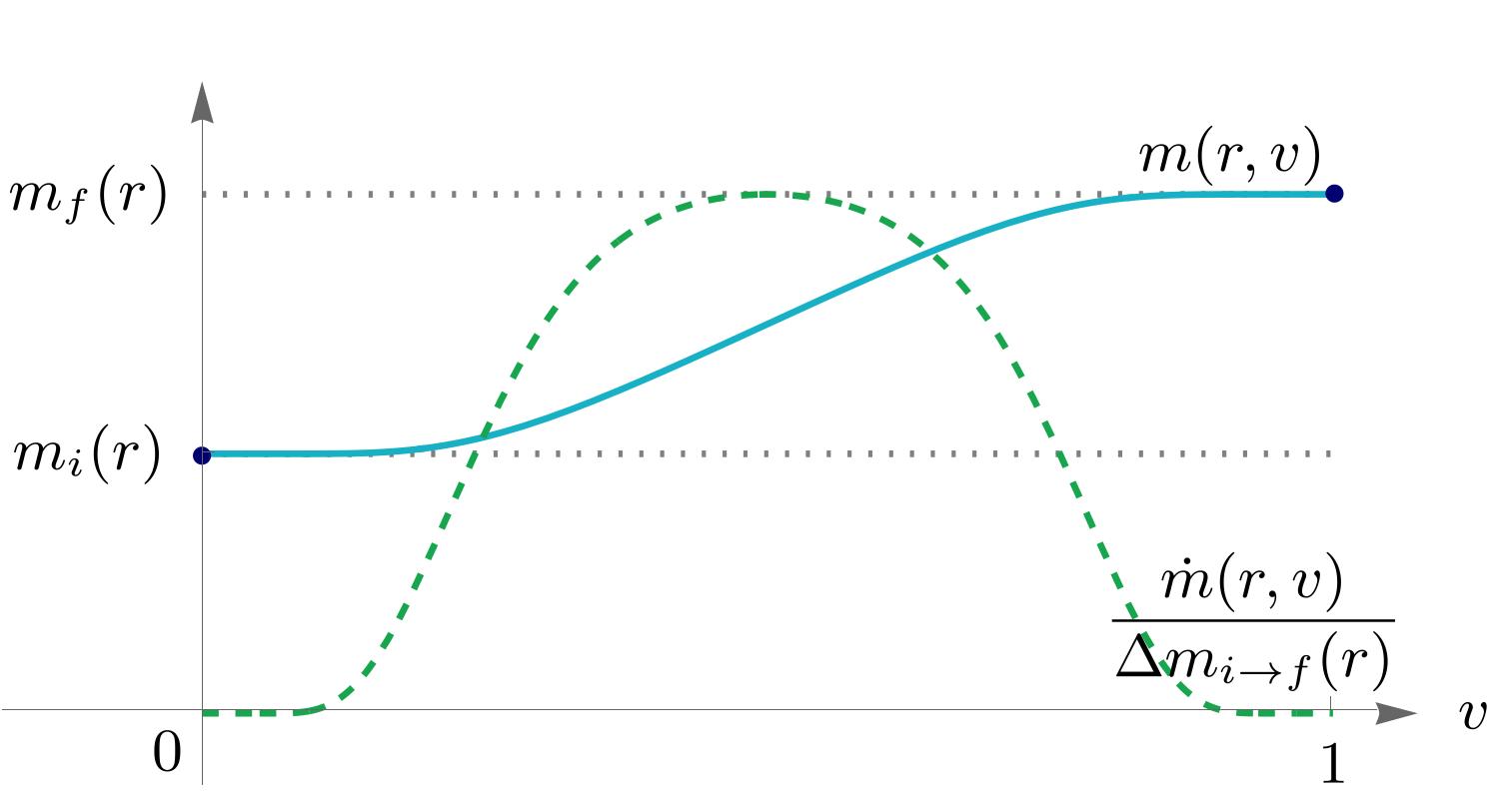}
	\caption{\label{Fig:InterpolatingFunctions} Left: Transition function $\sigma(v)$ (blue) defined in~\eqref{eq:Sigma} and its time derivative $\dot{\sigma}(v)$ (green dashed). Right: Interpolating mass function $m(r,v)$ (blue) defined in~\eqref{eq:MassFunctionInterpolation} and its time derivative $\dot{m}(r,v)$ (green dashed) for an example with $m_i(r)<m_f(r)=2m_i(r)$.}
\end{figure}

Consider an initial static geometry $(i)$ at time $v_i $ and a final static geometry $(f)$ at time $v_f $ described by mass functions $m_{i,f}(r)$ and radial functions $R_{i,f}(r)$, possibly with implicit dependencies on some internal scale parameters $l_{i,f}$. Our goal is to define functions $m(r,v)$ and $R(r,v)$ interpolating in time between the two geometries. To that end we introduce an infinitely differentiable smooth transition function
\be\label{eq:Sigma}
\sigma(v) = \frac{\tau(v)}{\tau(v) +\tau(1-v) } \,\,\,\quad \text{where} \quad \,\,\, \tau(v) = 
\begin{cases}
	e^{-\frac{1}{v}} & \text{for } v > 0\,,\\
	0 & \text{for } v \leq 0\,.
\end{cases}
\ee
This function satisfies $\sigma(v)=0$ for $v\leq 0$ and $\sigma(v)=1$ for $v\geq 1$ and $\dot{\sigma}(v)>0$ for $v \in (0,1)$, as shown in the left plot of~Fig.~\ref{Fig:InterpolatingFunctions}. (While this specific choice of interpolating function is needed to quantitatively produce the figures below, it should be clear that any monotonic interpolating function with the same asymptotics will lead to qualitatively similar results.)\\

\noindent {\it Interpolating mass function:} The interpolating mass function $m(r,v)$ should satisfy
\be\label{eq:Conditionsm}
\lim_{v\to v_{i,f}} m(r,v) = m_{i,f}(r)\,\,\,\quad \text{and} \quad \,\,\, \lim_{r\to \infty}\dot{m}(r,v)=0\,.
\ee
The last condition can be understood as demanding that the total ADM mass measured at infinity is preserved during the transition. Therefore, if $m_{i,f}(r) \to M_{i,f}$ for $r \to \infty$, we demand that $M_i = M_f \equiv M$. We can now use the function $\sigma(v)$ in~\eqref{eq:Sigma} to switch between the mass function $m_i(r)$ at time $v_i\equiv 0$ and the mass function $m_f(r)$ at time $v_f\equiv 1$ by defining 
\be\label{eq:MassFunctionInterpolation}
m(r,v) = \qty(1-\sigma(v)) m_i(r) + \sigma(v) m_f(r)\,,
\ee
as illustrated in the right plot of Fig.~\ref{Fig:InterpolatingFunctions}. 

Note that the time derivative of this mass function is
\be
\dot{m}(r,v) = \qty(m_f(r) - m_i(r))\dot{\sigma}(v) \equiv \Delta m_{i\to f}(r )\dot{\sigma}(v)\,.
\ee
Therefore $ \dot{m}(r,v) \to \qty(M_i - M_f)\dot{\sigma} (v)  = 0$ for $r\to \infty$ and hence  the last condition in~\eqref{eq:Conditionsm} is satisfied. Since $\dot{\sigma}>0$ during the transition, for geometries satisfying the assumptions of Section~\ref{SecSub:Rr}, i.e., for which $R(r,v)=r$, the NCC according to equation~\eqref{eq:NCClRr} is violated if 
\be\label{eq:NCCDeltaM}
\Delta m_{i\to f}(r ) < 0 \,\,\,\Leftrightarrow \,\,\,m_f(r) < m_i(r)\,.
\ee 
We will provide examples for this NCC violation in transitions from singular black holes to regular black holes in Section~\ref{Sec:SBHtoRBH}.

\noindent Now, if the functional form of the mass profile $m(r)$ for the initial and final geometries is identical up to a length parameter $l_{i,f}$, we may alternatively define an interpolating mass function $m(r,v)$ by making use of an interpolating length function $l^n(v)$,
\be\label{eq:LengthFunctionInterpolation}
l^n(v) = \qty(1- \sigma(v))l^n_i + \sigma(v) l^n_f\,,
\ee
where $n$ is the power with which the parameter $l_{i,f}$ appears in the definition of the mass function $m(r)$ for the static initial and final geometry (e.g.~later $n=2$ for Bardeen and $n=3$ for Dymnikova). The function $l^n(v)$ is then taken to be part of what defines the interpolating mass function by
\be\label{eq:MassFunctionInterpolation2}
m(r,v) = m\qty(r;l^n =l^n(v))\,.
\ee
Note that by the chain rule the time derivative of this mass function is
\be\label{eq:MassFunctionInterpolation2Dot}
\dot{m}(r,v) = \qty(\partial_{l^n} m)\partial_v \qty(l^n(v)) = \qty(\partial_{l^n} m) \qty(l_f^n - l_i^n) \dot{\sigma}(v) \equiv \qty(\partial_{l^n} m)  \dot{\sigma}(v) \Delta l^n_{i\to f} \,.
\ee
Therefore since $\dot{\sigma}>0$, and provided $\qty(\partial_{l^n} m) < 0$, for geometries satisfying the assumptions of Section~\ref{SecSub:Rr}, i.e., for which $R(r,v)=r$, the NCC according to equation~\eqref{eq:NCClRr} is violated if 
\be\label{eq:NCCDeltaL}
 \Delta l^n_{i\to f} > 0 \,\,\,\Longleftrightarrow \,\,\,l^n_f > l^n_i\,.
\ee 
We will provide examples for this NCC violation in transitions from regular black holes to horizonless compact objects in Section~\ref{Sec:RBHtoUCO}.
\\

\noindent {\it Interpolating radial function:} We can repeat the above constructions to define an interpolating radial function $R(r,v)$ such that
\be\label{eq:ConditionsR}
\lim_{v\to v_{i,f}} R(r,v) = R_{i,f}(r)\,\,\,\quad \text{and} \quad \,\,\, \lim_{r\to \infty}R(r,v) \simeq r\,.
\ee
The last condition imposes that at infinity the angular part of the line element~\eqref{eq:MetricImplodingSphericalSymmetry} reduces to the usual area element for a 2-sphere of radius $r$. When combined with the last condition in~\eqref{eq:Conditionsm} we therefore demand that at large distances the metric before, during and after the transition is well approximated by the asymptotically flat Schwarzschild metric with mass $M$. Using again the function $\sigma(v)$ in~\eqref{eq:Sigma} to switch between the radial function $R_i(r)$ at time $v_i\equiv 0$ and the radial function $R_f(r)$ at time $v_f\equiv 1$, we define
\be\label{eq:RFunctionInterpolation}
R(r,v) = \qty(1-\sigma(v))\; R_i(r) + \sigma(v)\; R_f(r)\,.
\ee
Note that the second radial derivative of this function is
\be
R''(r,v) = \qty(1-\sigma(v))\; R_i''(r) + \sigma(v)\; R_f''(r)\,.
\ee
Now if the initial geometry satisfies the assumptions of Section~\ref{SecSub:Rr}, i.e., $R_i(r)=r$, then $R''_i = 0$ and therefore, using that $\dot{\sigma}>0$, the NCC according to equation~\eqref{eq:NCCk} will be violated if $R_f'' > 0$. We will provide examples for this NCC violation in transitions from singular black holes to bouncing or wormhole geometries in Section~\ref{Sec:SBHtoBounce}.\\

\noindent{Finally, if the functional form of the radial function $R(r)$ for the initial and final geometries is identical up a length parameter $l_{i,f}$, we may alternatively define an interpolating length function $R(r,v)$ by making use of the interpolating length function $l^n(v)$ in~\eqref{eq:LengthFunctionInterpolation}. Hence in this case we will define
\be\label{eq:RFunctionInterpolation2}
R(r,v) = R\qty(r;l^n =l^n(v))\,.
\ee
Note that the second radial derivative of this function is simply
\be\label{eq:NCCRconcrete}
R''(r,v) = R''(r,l^n(v))\,.
\ee
Therefore, if the initial and final geometries have $R''(r)>0$ the NCC will be be violated according to equation~\eqref{eq:NCCk}, independent of the value of the length parameters. We will provide an example for this NCC violation in a transition from a black bounce to a naked wormhole in Section~\ref{Sec:SVBHtoSVWormhole}.
}\\

\subsection{Transitions from singular to regular black holes}\label{Sec:SBHtoRBH}

In the following subsubsections we illustrate the violation of the NCC in transitions from a singular to a regular black hole with $R(r,v)=r$ according to equation~\eqref{eq:NCClRr} using the interpolating mass function~\eqref{eq:MassFunctionInterpolation}. The quantities of interest are the metric function $f(r,v)$ defined in~\eqref{eq:MetricImplodingSphericalSymmetry} and the left hand side of the NCC in equation~\eqref{eq:NCClRr},
\be\label{eq:fNCC}
f(r,v) = 1 - \frac{2 m(r,v)}{r}\,\,\,\quad \text{and} \quad \,\,\, R_{\mu\nu}\; l^\mu l^\nu = \frac{2 \dot{m}(r,v)}{r^2}\,.
\ee
As the initial geometry we first choose a  black hole with an integrable singularity for which the metric function $f(r,v)$ is finite at $r=0$ due to a positive length parameter $l_i$.  We will also consider the limit $l_i \to 0$ corresponding to a Schwarzschild initial geometry. As the final geometries we choose two types of regular black holes with smooth de Sitter cores and length parameters denoted by $l_f$.\\

\subsubsection{BH with integrable or Schwarzschild singularity $\longrightarrow$ Bardeen RBH}

\begin{figure}[h!]
	\centering
	\includegraphics[width=.45\textwidth]{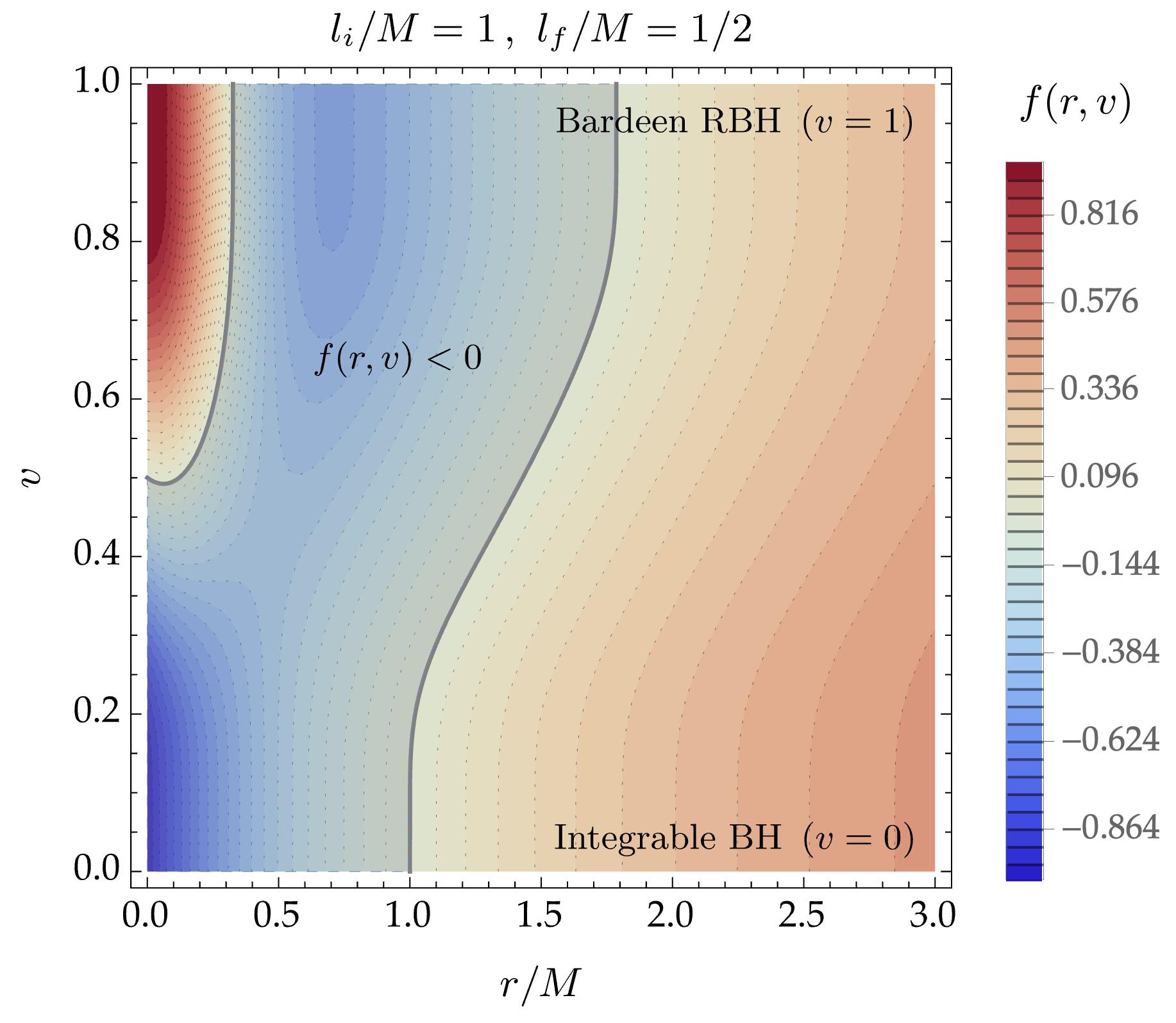}
	\hspace{0.1cm}
	\includegraphics[width=.45\textwidth]{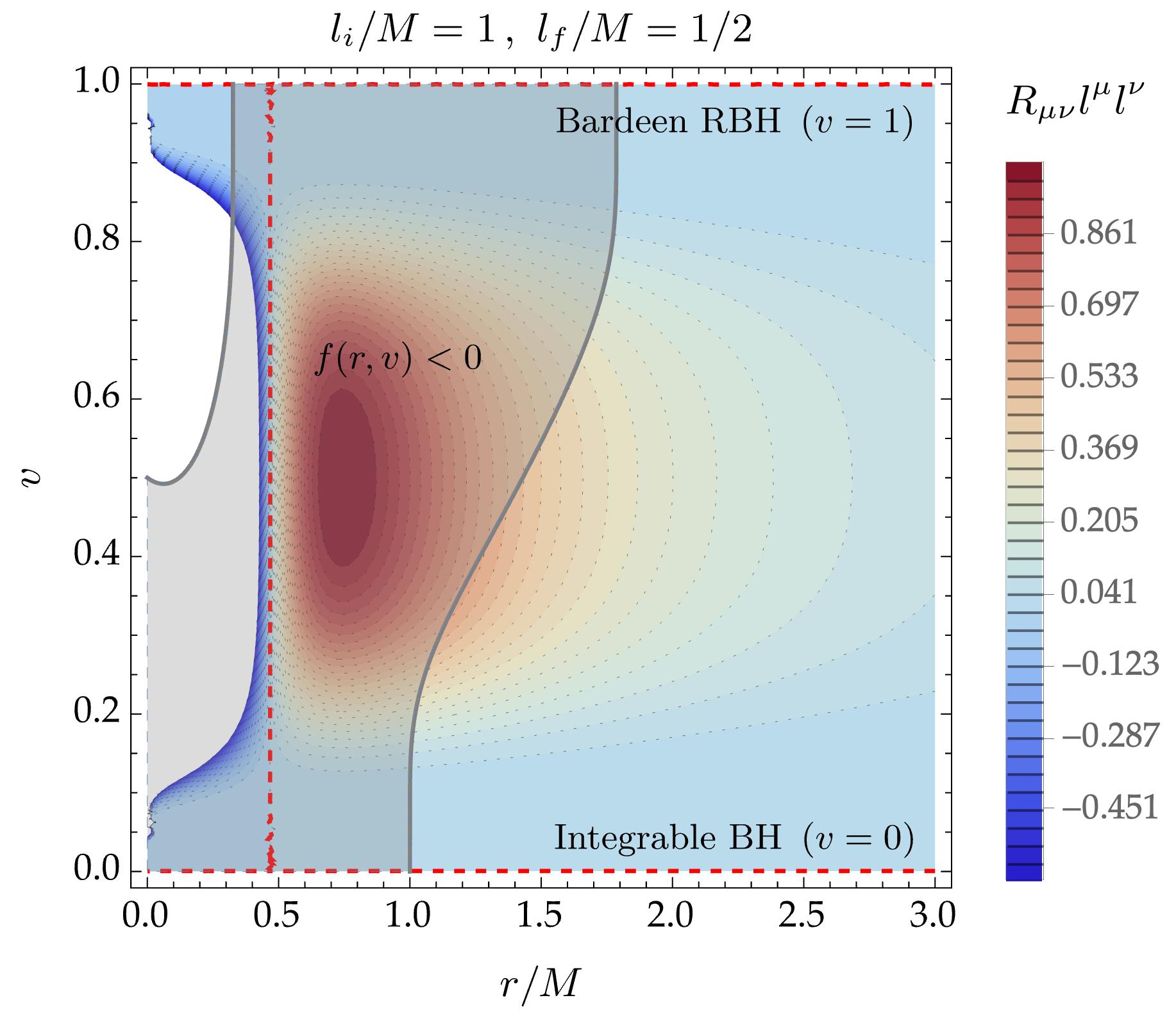}
	\caption{\label{Fig:BHIntegrableToRBHBardeen} Plots of the metric function $f(r,v)$ (left) and the NCC (right) in~\eqref{eq:fNCC} for a transition from a black hole with integrable singularity to a Bardeen regular black hole with mass functions~\eqref{eq:MassFunctionsBHIntegrableBardeen} via the interpolating function $m(r,v)$ defined in~\eqref{eq:MassFunctionInterpolation}. Along the red dashed lines $R_{\mu\nu}\;l^\mu l^\nu = 0$, marking the edge of the NCC-violating region. Note that at any constant-$v$ slice the intersection with the vertical red dashed line is defined by the root of $\Delta m_{i\to f}(r)=0$ and therefore of $m_i(r)=m_f(r)$, according to equation~\eqref{eq:NCCDeltaM}. The gray-shaded region marks the trapped region where $f(r,v)<0$. The uncolored region in the right plot is bounded by the minimum value of the legend and represents a region of extremely high and off-scale NCC violation. The length parameters are set to $l_i/M=1$ and $l_f/M=1/2$.}
\end{figure}

Fig.~\ref{Fig:BHIntegrableToRBHBardeen} shows the metric function $f(r,v)$ and left hand side of the NCC in equation~\eqref{eq:fNCC} for an initial geometry corresponding to a black hole with an integrable singularity and a final geometry corresponding to a regular Bardeen black hole~\cite{Bardeen:1968bh}. The mass functions are given by
\be\label{eq:MassFunctionsBHIntegrableBardeen}
m_{i}(r) = M\qty(\frac{r}{r+l_i}) \,\,\,\quad \text{and} \quad \,\,\, m_{f}(r) = M\qty(\frac{r^3}{\qty(r^2 + l_f^2)^{\frac{3}{2}}})\,.
\ee
It is easy to check that $\left(2 m' - r m''\right)>0$ {(that is, $\rho' < 0$)} for these specific mass profiles, so (as per our previous discussion) for both the initial and final spacetimes the NCC is satisfied in all null directions, not just the radial null directions. 

\begin{figure}[h!]
	\centering
	\includegraphics[width=.45\textwidth]{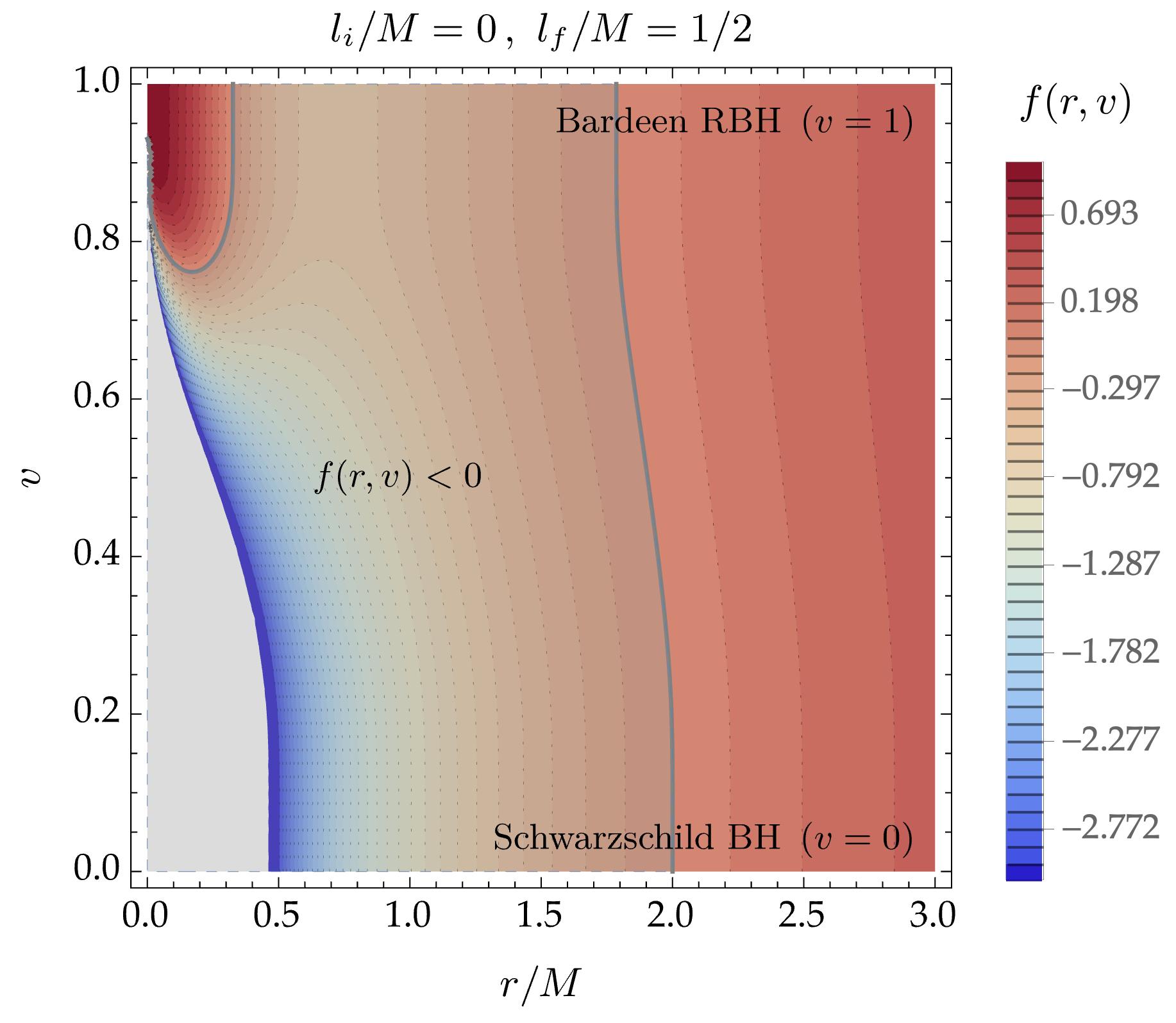}
	\hspace{0.1cm}
	\includegraphics[width=.45\textwidth]{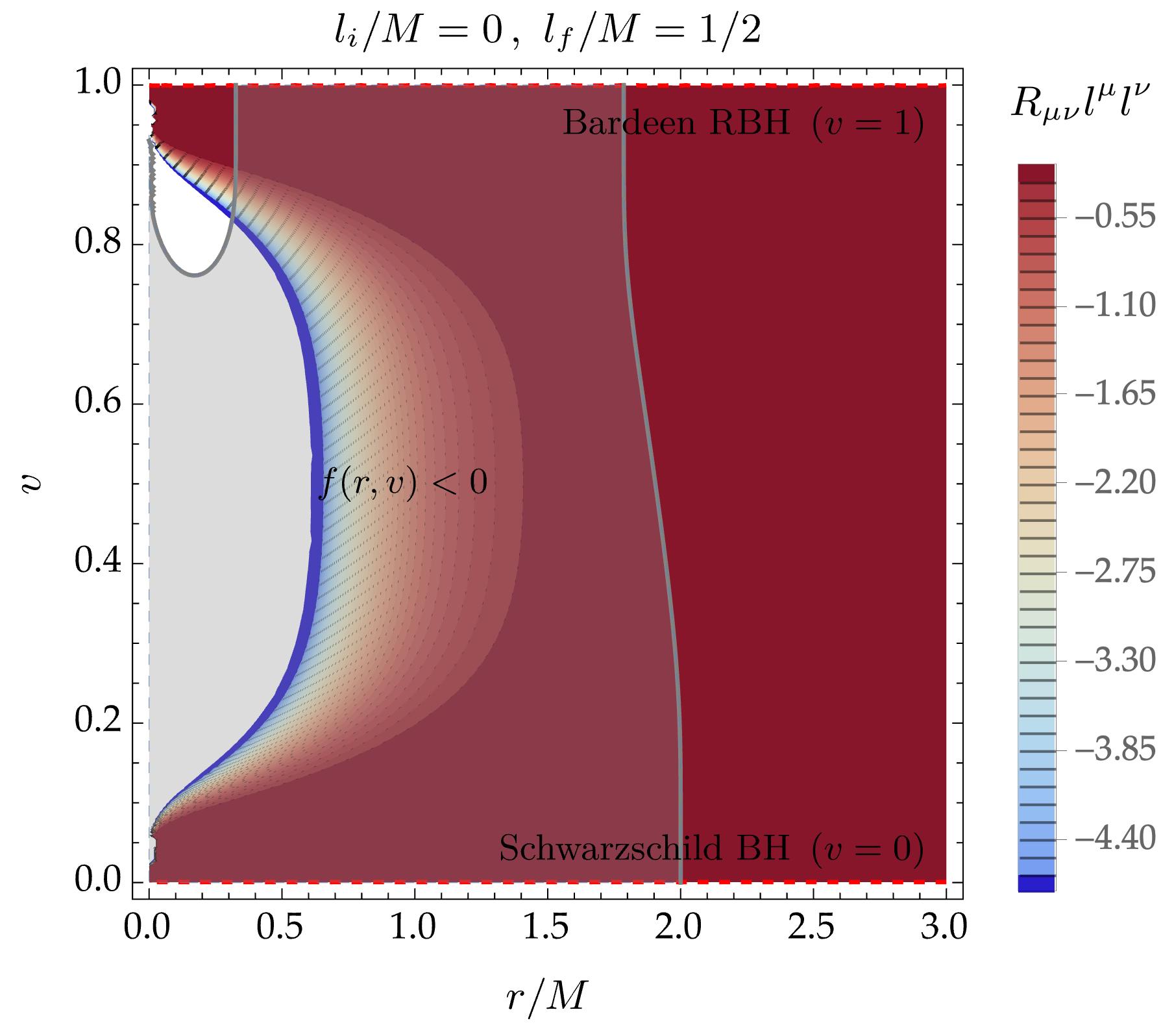}
	\caption{\label{Fig:BHSchwarzschildToRBHBardeen} Plots of the metric function $f(r,v)$ (left) and the NCC (right) in~\eqref{eq:fNCC} for a transition from a Schwarzschild black hole to a Bardeen regular black hole with mass functions~\eqref{eq:MassFunctionsBHIntegrableBardeen} via the interpolating function $m(r,v)$ defined in~\eqref{eq:MassFunctionInterpolation}. Along the red dashed horizontal lines $R_{\mu\nu}l^\mu l^\nu = 0$, and the NCC is satisfied. Everywhere else the NCC is violated in accordance with equation~\eqref{eq:NCCDeltaM}, as $m_f(r)<m_i(r)$ at any intermediate $v$. The gray-shaded region marks the trapped region where $f(r,v)<0$. The uncolored region in the right plot is bounded by the minimum value of the legend and represents a region of extremely high and off-scale NCC violation. The length parameters are set to $l_i/M=0$ and $l_f/M=1/2$.}
\end{figure}

Fig.~\ref{Fig:BHSchwarzschildToRBHBardeen} shows the same quantities for $l_i \to 0$ when the initial geometry is Schwarzschild.

\enlargethispage{120pt}

\newpage

\subsubsection{BH with integrable or Schwarzschild singularity $\longrightarrow$ Dymnikova RBH}

\begin{figure}[htb!]
	\centering
	\includegraphics[width=.45\textwidth]{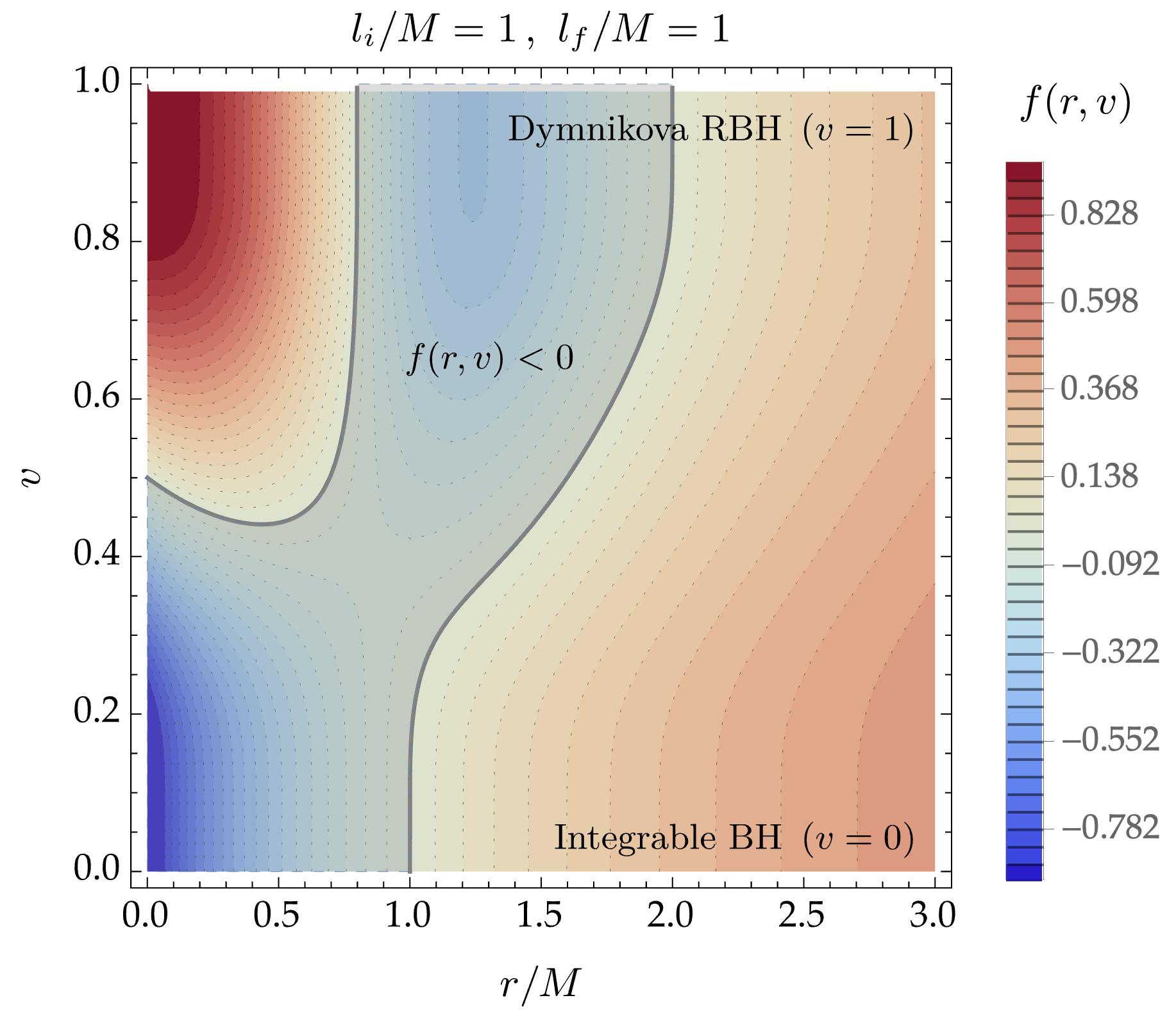}
	\hspace{0.1cm}
	\includegraphics[width=.45\textwidth]{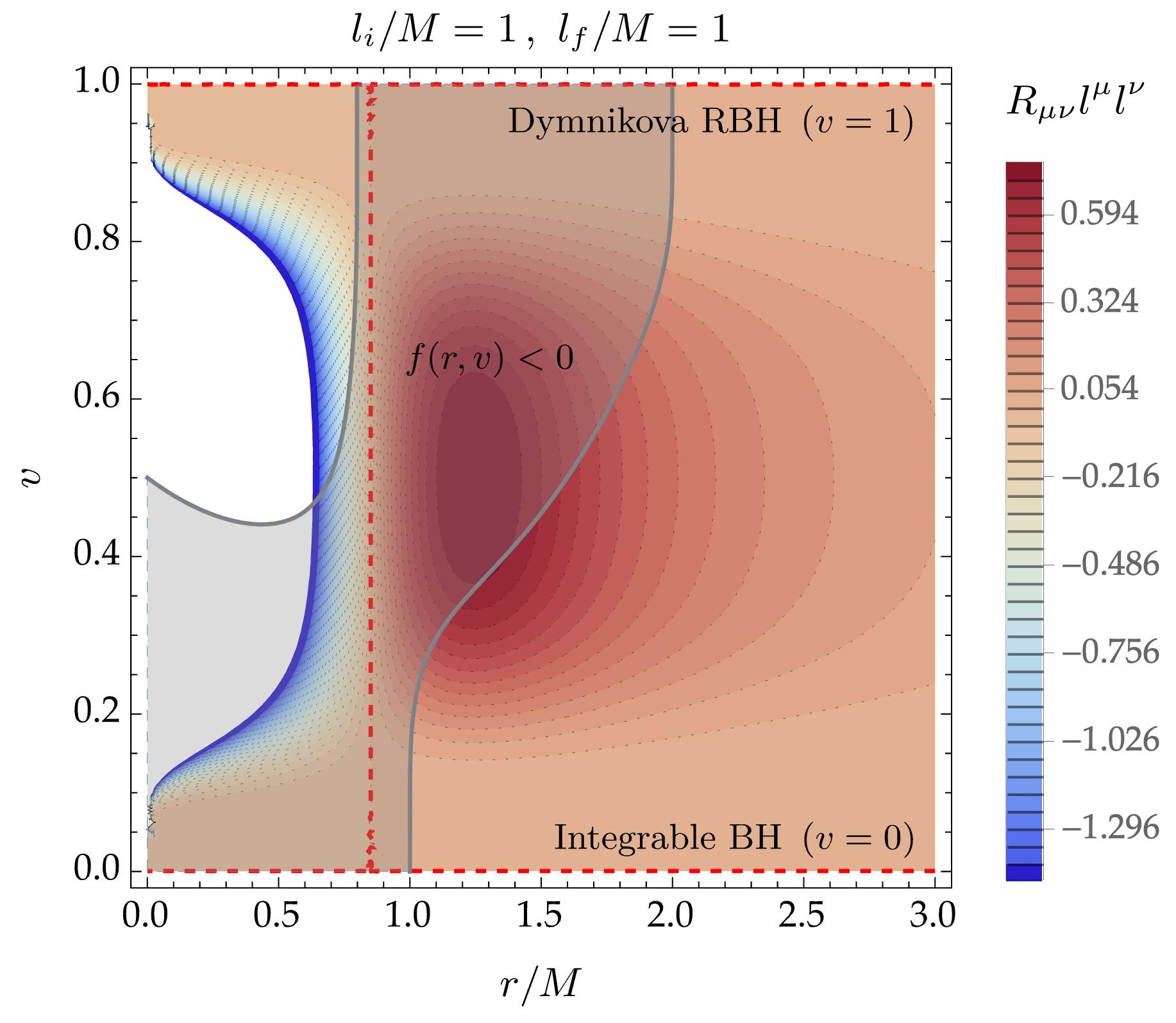}
	\caption{\label{Fig:BHIntegrableToRBHDymnikova} Plots of the metric function $f(r,v)$ (left) and the NCC (right) in~\eqref{eq:fNCC} for a transition from a black hole with integrable singularity to a Dymnikova regular black hole with mass functions~\eqref{eq:MassFunctionsBHIntegrableDymnikova} via the interpolating function $m(r,v)$ defined in~\eqref{eq:MassFunctionInterpolation}. Along the red dashed lines $R_{\mu\nu}\; l^\mu l^\nu = 0$, marking the edge of the NCC-violating region. Note that at any constant-$v$ slice the intersection with the vertical red dashed line is defined by the root of $\Delta m_{i\to f}(r)=0$ and therefore of $m_i(r)=m_f(r)$, according to equation~\eqref{eq:NCCDeltaM}. The gray-shaded region marks the trapped region where $f(r,v)<0$. The uncolored region in the right plot is bounded by the minimum value of the legend and represents a region of extremely high and off-scale NCC violation. The length parameters are set to $l_i/M=1$ and $l_f/M=1$.}
\end{figure}

\begin{figure}[htb!]
	\centering
	\includegraphics[width=.45\textwidth]{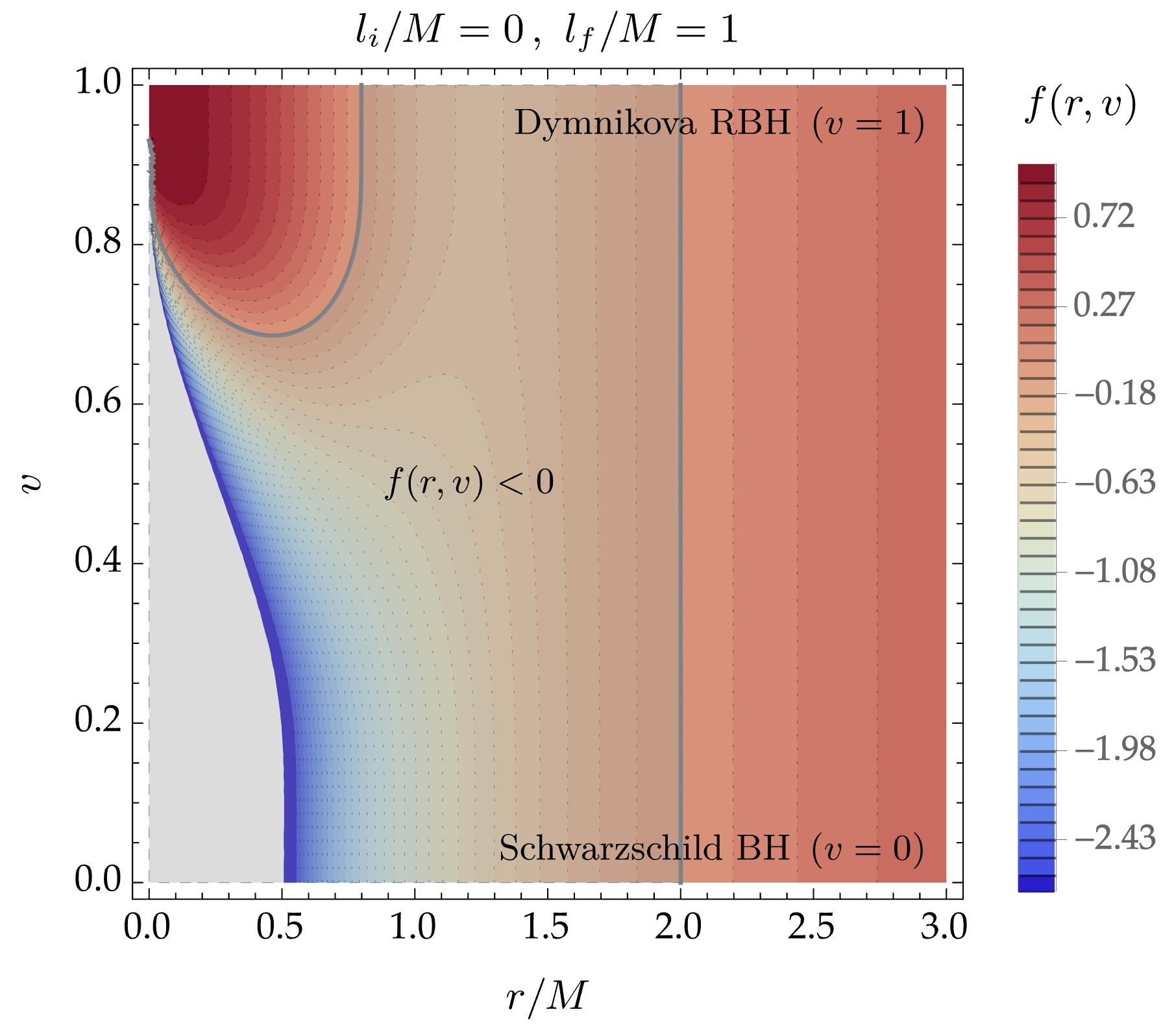}
	\hspace{0.1cm}
	\includegraphics[width=.45\textwidth]{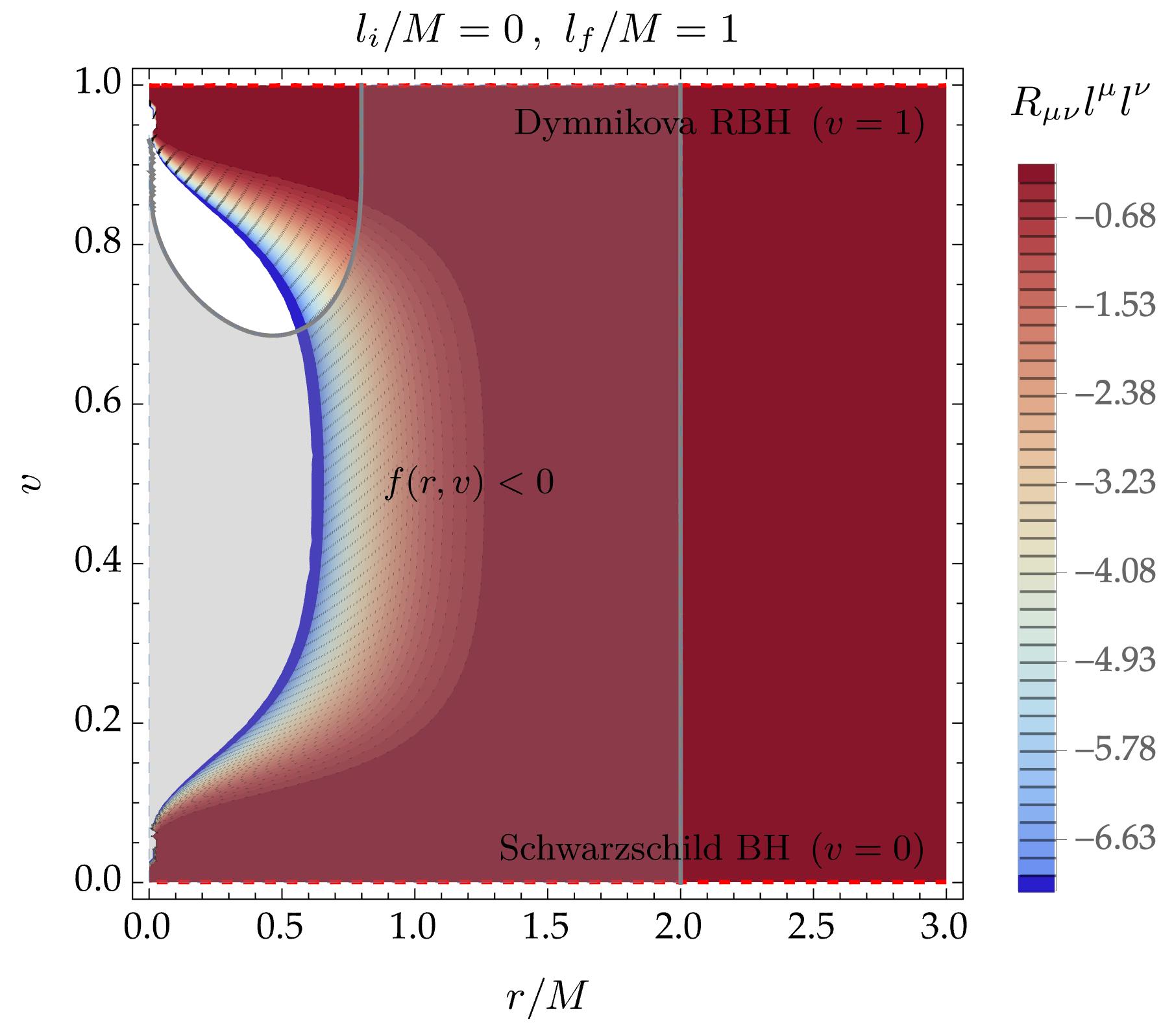}
	\caption{\label{Fig:BHSchwarzschildToRBHDymnikova} Plots of the metric function $f(r,v)$ (left) and the NCC (right) in~\eqref{eq:fNCC} for a transition from a Schwarzschild black hole to a Dymnikova regular black hole with mass functions~\eqref{eq:MassFunctionsBHIntegrableDymnikova} via the interpolating function $m(r,v)$ defined in~\eqref{eq:MassFunctionInterpolation}. Along the horizontal red dashed lines $R_{\mu\nu}l^\mu l^\nu = 0$, and the NCC is satisfied. Everywhere else the NCC is violated {in accordance with equation~\eqref{eq:NCCDeltaM}, as $m_f(r)<m_i(r)$ at any intermediate $v$}. The gray-shaded region marks the trapped region where $f(r,v)<0$. The uncolored region in the right plot is bounded by the minimum value of the legend and represents a region of extremely high and off-scale NCC violation.
    The length parameters are set to $l_i/M=0$ and $l_f/M=1$.}
\end{figure}

Fig.~\ref{Fig:BHIntegrableToRBHDymnikova} shows the metric function $f(r,v)$ and left hand side of the NCC in equation~\eqref{eq:fNCC} for an initial geometry corresponding to a black hole with an integrable singularity and a final geometry corresponding to a regular Dymnikova black hole~\cite{Dymnikova:1992ux}. The mass functions are given by
\be\label{eq:MassFunctionsBHIntegrableDymnikova}
m_{i}(r) = M\qty(\frac{r}{r+l_i}) \,\,\,\quad \text{and} \quad \,\,\, m_{f}(r) = M\qty(1-e^{-\frac{r^3}{l_f^3}})\,.
\ee
It is easy to check that $\left(2 m' - r m''\right)>0$ {(that is, $\rho'<0$)} for these specific mass profiles, so (as per our previous discussion) for both the initial and final spacetimes the NCC is satisfied in all null directions, not just the radial null directions. Fig.~\ref{Fig:BHSchwarzschildToRBHDymnikova} instead shows the same quantities for $l_i \to 0$ when the initial geometry is Schwarzschild. 

The same qualitative features for the transition to a Dymnikova black hole can be observed as for the transition to a Bardeen black hole, which shows the insensitivity to the particular regular black hole final state.


\clearpage
\subsection{Transitions from regular black holes to horizonless compact objects}\label{Sec:RBHtoUCO}

In the following we provide examples for the violation of the NCC according to equation~\eqref{eq:NCClRr} in transitions from an initial geometry describing a regular black hole with length parameter $l_i$ to a final geometry of the same type but with different value of this length parameter such that the spacetime describes a horizonless compact object. We denote the final value of the length parameter by $l_f$. To that end we {now make use of the mass function~\eqref{eq:MassFunctionInterpolation2} which uses the interpolating length function~\eqref{eq:LengthFunctionInterpolation}.}

\subsubsection{Bardeen RBH $\longrightarrow$ Bardeen HCO}

\begin{figure}[!h]
	\centering
	\includegraphics[width=.45\textwidth]{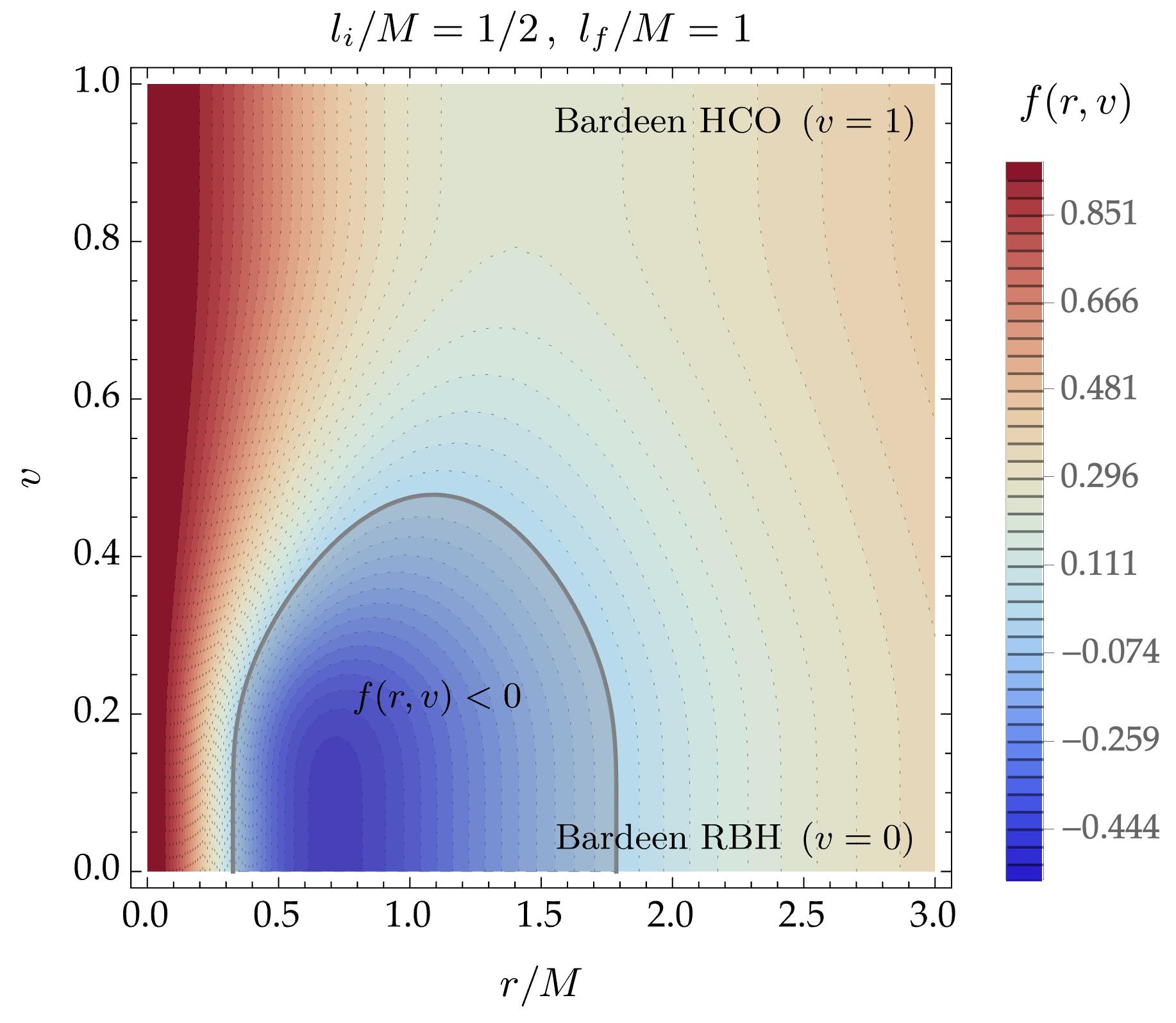}
	\hspace{0.1cm}
	\includegraphics[width=.45\textwidth]{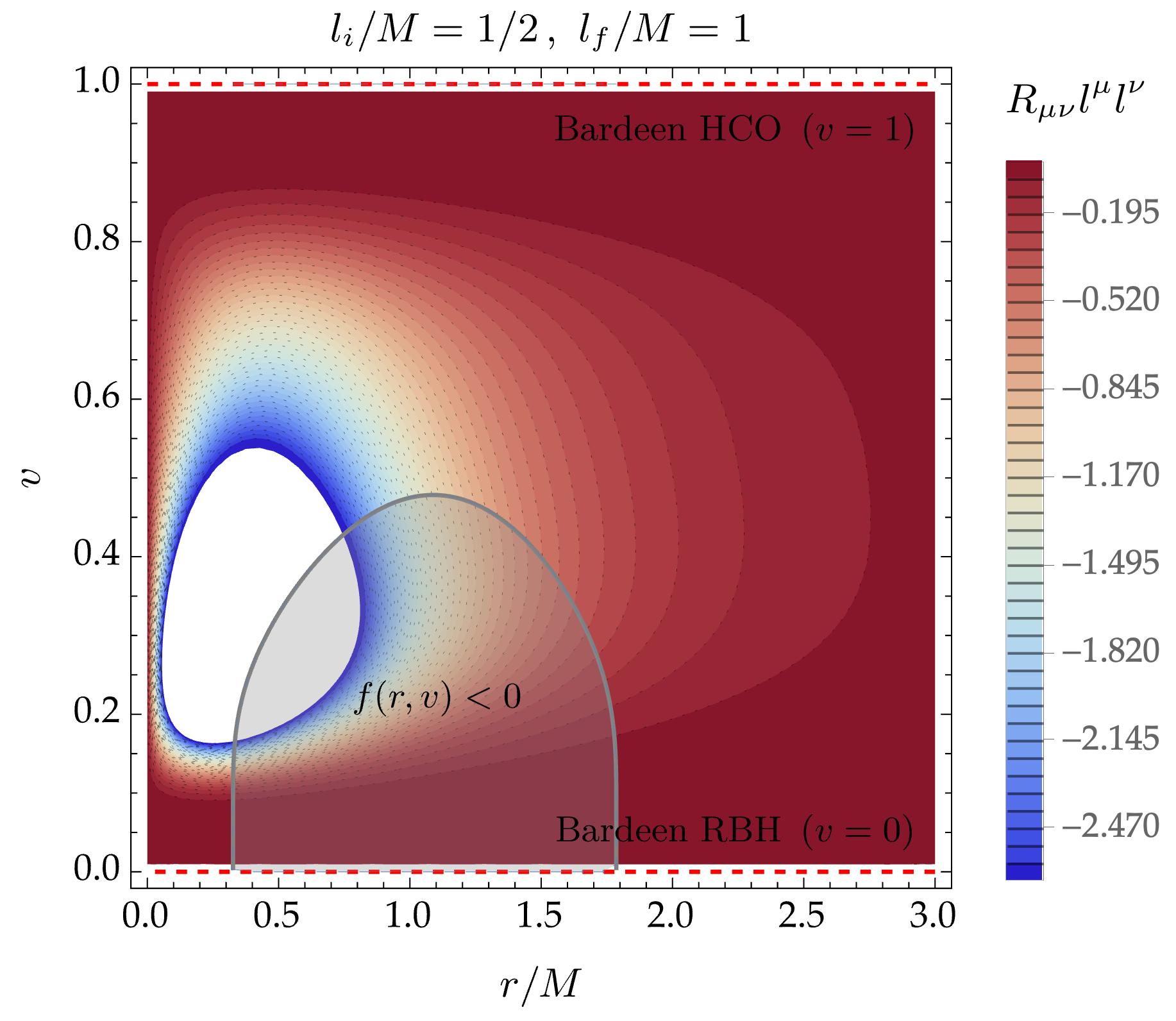}
	\caption{\label{Fig:RBHBardeenToUCOBardeen} Plots of the metric function $f(r,v)$ (left) and the NCC (right) in~\eqref{eq:fNCC} for a transition from a Bardeen regular black hole to a Bardeen horizonless compact object with interpolating mass function~\eqref{eq:MassFunctionsRBHBardeenUCOBardeen}. The gray-shaded region marks the trapped region where $f(r,v)<0$. {The uncolored region in the right plot is bounded by the minimum value of the legend and represents a region of extremely high and off-scale NCC violation. {Along the red dashed horizontal lines $R_{\mu\nu}l^\mu l^\nu = 0$, and the NCC is satisfied.} {Everywhere else in between the initial and final geometries, the NCC is violated, in accordance with equations~\eqref{eq:MassFunctionInterpolation2Dot} and~\eqref{eq:NCCDeltaL}, as $\qty(\partial_{l^2} m) < 0$ and $l_f>l_i$.}}
    The length parameters are set to $l_i/M=1/2$ and $l_f/M=1$.}
\end{figure}

For a transition from a regular black hole to {a horizonless compact object} based on a Bardeen geometry, we set the mass function to be
\be\label{eq:MassFunctionsRBHBardeenUCOBardeen}
m(r,v) =M\qty(\frac{r^3}{\qty(r^2 + l^2(v))^{\frac{3}{2}}})\,.
\ee
Fig.~\ref{Fig:RBHBardeenToUCOBardeen} shows the metric function $f(r,v)$ and left hand side of the NCC in equation~\eqref{eq:fNCC} for an initial geometry corresponding to a Bardeen regular black hole and a final geometry corresponding to a Bardeen horizonless compact object.
Note that both initial and final spacetimes satisfy the NCC, while the interpolating spacetime violates the NCC.

\subsubsection{Dymnikova RBH $\longrightarrow$ Dymnikova HCO}

\begin{figure}[!h]
	\centering
	\includegraphics[width=.45\textwidth]{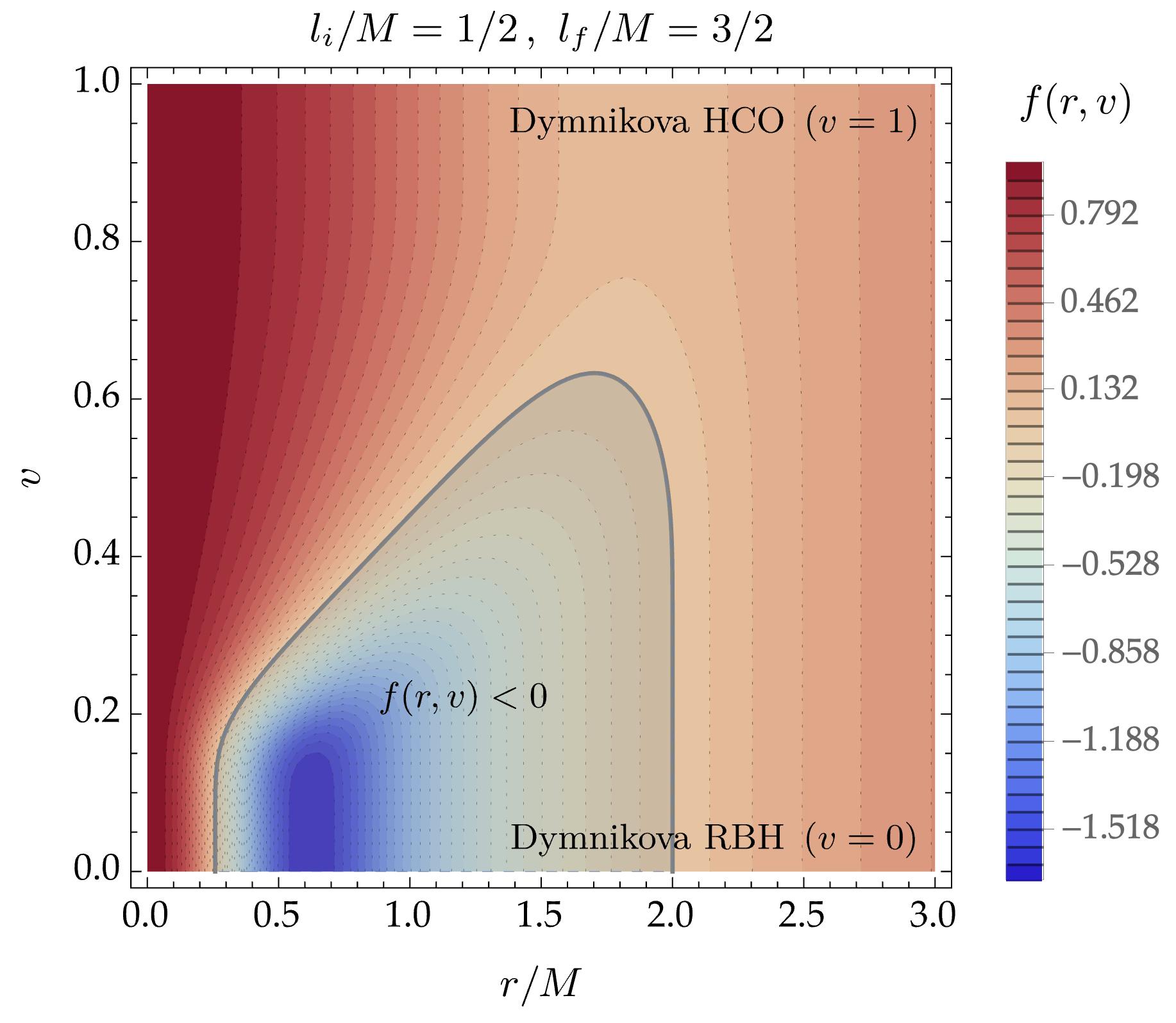}
	\hspace{0.1cm}
	\includegraphics[width=.45\textwidth]{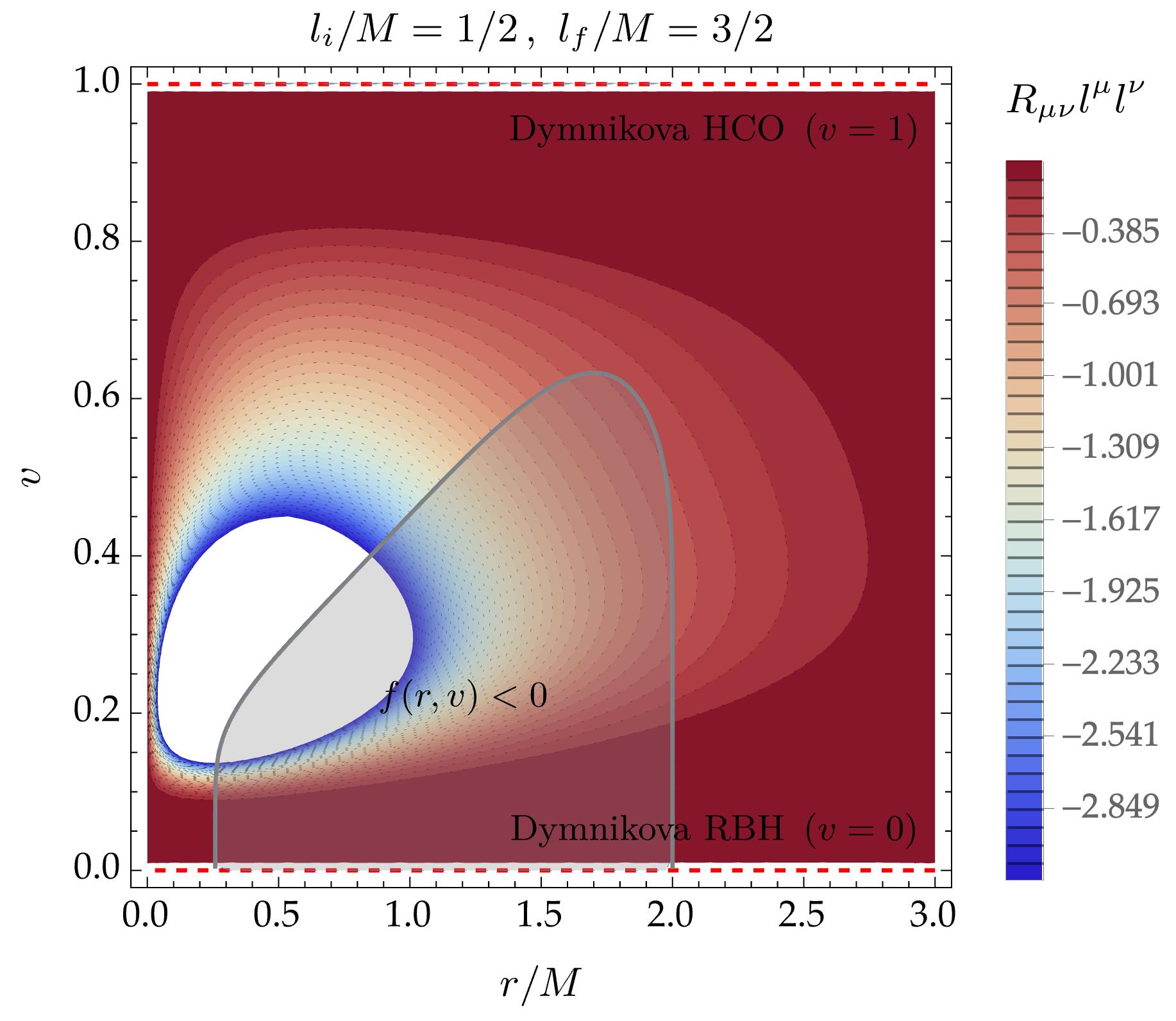}
	\caption{\label{Fig:RBHDymnikovaToUCODymnikova} Plots of the metric function $f(r,v)$ (left) and the NCC (right) in~\eqref{eq:fNCC} for a transition from a Dymnikova regular black hole to a Dymnikova horizonless compact object with interpolating mass function~\eqref{eq:MassFunctionsRBHDymnikovaUCODymnikova}. The gray-shaded region marks the trapped region where $f(r,v)<0$. {The uncolored region in the right plot is bounded by the minimum value of the legend and represents a region of extremely high and off-scale NCC violation.} {Along the red dashed horizontal lines $R_{\mu\nu}l^\mu l^\nu = 0$, and the NCC is satisfied.} {Everywhere else in between the initial and final geometries, the NCC is violated, in accordance with  equations~\eqref{eq:MassFunctionInterpolation2Dot} and~\eqref{eq:NCCDeltaL}, as $\qty(\partial_{l^3} m) < 0$ and $l_f>l_i$.} 
    The length parameters are set to $l_i/M=1/2$ and $l_f/M=3/2$.}
\end{figure}

For a transition from a regular black hole to a horizonless compact object based on a Dymnikova geometry, we set the mass function to be
\be\label{eq:MassFunctionsRBHDymnikovaUCODymnikova}
m(r,v) = M\qty(1-e^{-\frac{r^3}{l_f^3(v)}})\,.
\ee
Fig.~\ref{Fig:RBHDymnikovaToUCODymnikova} shows the metric function $f(r,v)$ and the NCC in equation~\eqref{eq:fNCC} for an initial geometry corresponding to a Dymnikova regular black hole and a final geometry corresponding to a Dymnikova horizonless compact object. Note that both initial and final spacetimes satisfy the NCC, while the interpolating spacetime violates the NCC. As before we observe the same qualitative features for the transition to a Dymnikova HCO as for the transition to a Bardeen HCO, which shows the insensitivity to the particular HCO final state.

\subsection{Transitions from singular black holes to bouncing or wormhole geometries}\label{Sec:SBHtoBounce}

In the following we provide examples for the violation of the NCC in transitions from a black hole with an integrable singularity satisfying the assumptions of Section~\ref{SecSub:Rr}, to a regular Simpson--Visser geometry~\cite{Simpson:2018tsi,Lobo:2020ffi,Franzin:2021vnj,Lobo:2020kxn} satisfying the assumptions of Section~\ref{SecSub:Rr0}, according to equation~\eqref{eq:NCCk}. We use the interpolating mass function $m(r,v)$ in~\eqref{eq:MassFunctionInterpolation} and radial function $R(r,v)$ in~\eqref{eq:RFunctionInterpolation}, and set
\be\label{eq:MassFunctionsBHSingularSV}
m_{i}(r) =  M\qty(\frac{r}{r+l_i})  \,\,\,\quad \text{and} \quad \,\,\, M_{f}(r) = M\,,
\ee
as well as
\be\label{eq:RadialFunctionsBHSingularSV}
R_{i}(r) = r \,\,\,\quad \text{and} \quad \,\,\, R_{f}(r) = \sqrt{r^2 + l_f^2}\,.
\ee
The quantities of interest are the metric function $f(r,v)$ defined in~\eqref{eq:MetricImplodingSphericalSymmetry} and the left hand side of the NCC in equation~\eqref{eq:NCCk},
\be\label{eq:fNCC2}
f(r,v) = 1 - \frac{2 m(r,v)}{R(r,v)}\,\,\,\quad \text{and} \quad \,\,\, R_{\mu\nu}k^\mu k^\nu = - \frac{2 R''(r,v)}{R(r,v)}\,.
\ee
Note that now the initial spacetime satisfies the NCC, while the final spacetime violates the NCC.
So it is now not too surprising that the interpolating spacetime also violates the NCC.

\subsubsection{BH with integrable or Schwarzschild singularity $\longrightarrow$  SV black bounce}

\begin{figure}[!h]
	\centering
	\includegraphics[width=.45\textwidth]{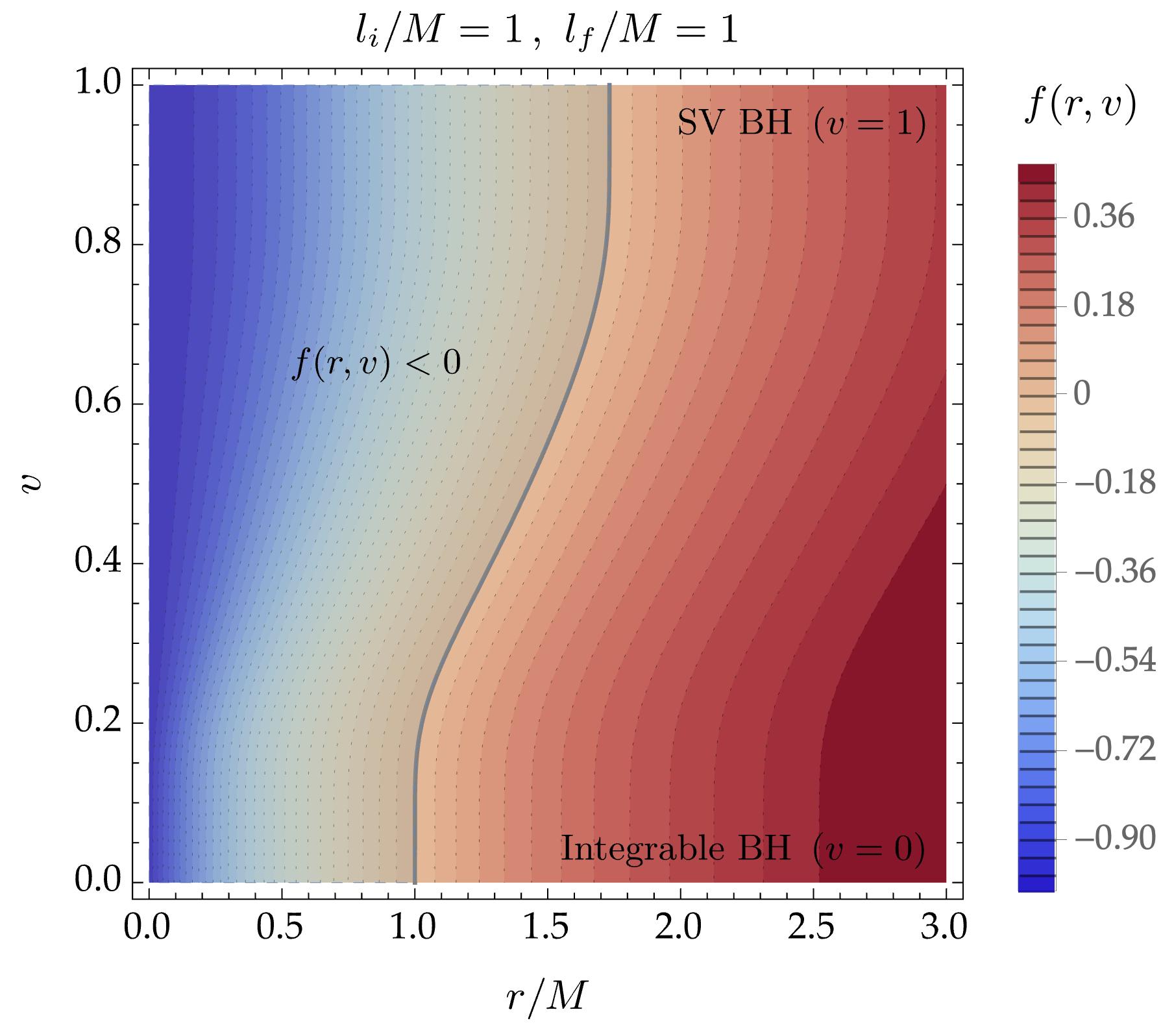}
	\hspace{0.1cm}
	\includegraphics[width=.45\textwidth]{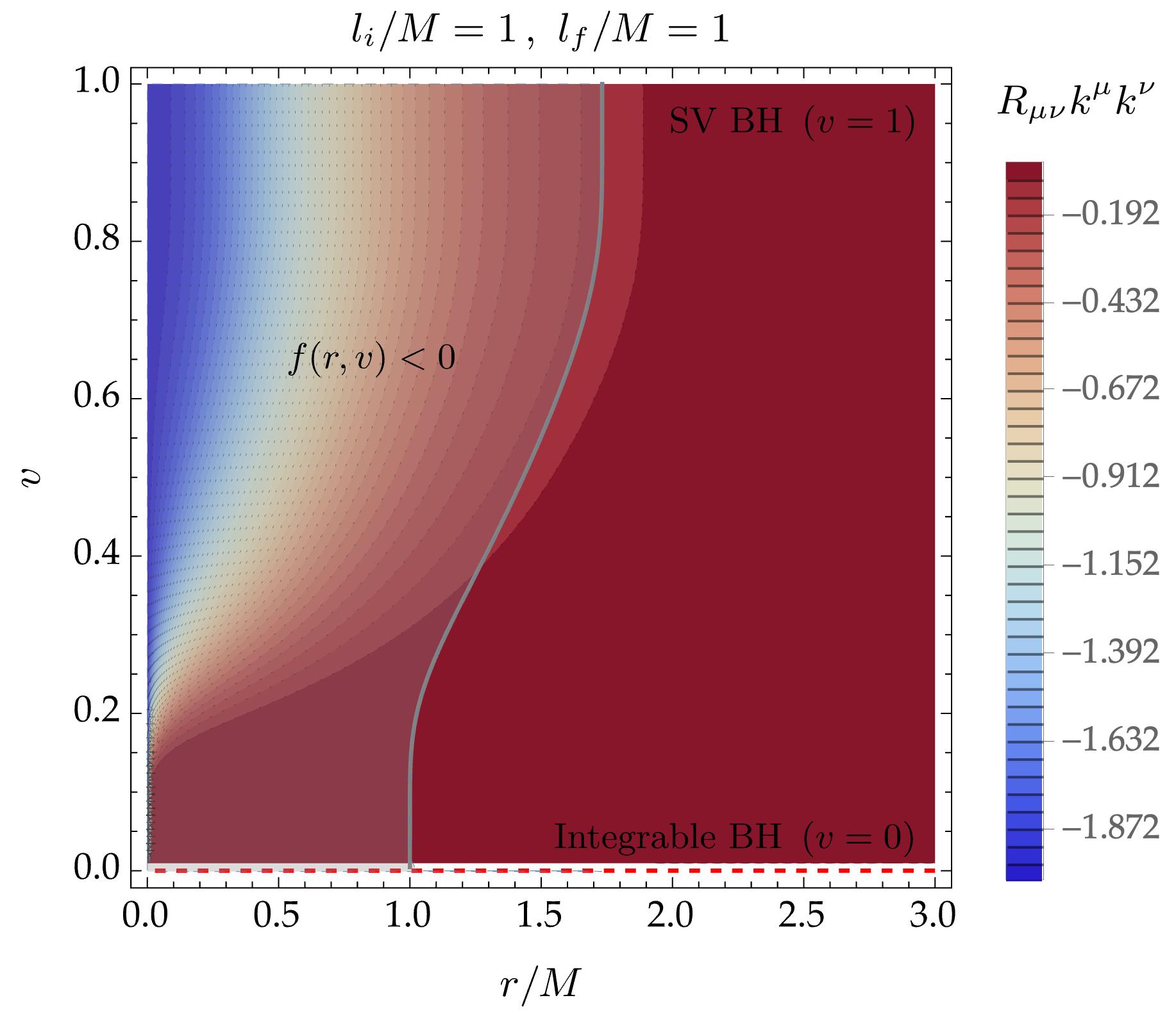}
	\caption{\label{Fig:BHIntegrableToSV} Plots of the metric function $f(r,v)$ (left) and the NCC (right) in~\eqref{eq:fNCC2} for a transition from a black hole with integrable singularity to a Simpson--Visser black hole with mass and radial functions~\eqref{eq:MassFunctionsBHSingularSV} and~\eqref{eq:RadialFunctionsBHSingularSV}, via the interpolating mass and radial functions~\eqref{eq:MassFunctionInterpolation} and~\eqref{eq:RFunctionInterpolation}. The gray-shaded region marks the trapped region where $f(r,v)<0$. {Along the red dashed horizontal line $R_{\mu\nu}k^\mu k^\nu = 0$, and the NCC is satisfied. Everywhere else the NCC is violated, in accordance with equation~\eqref{eq:NCCRconcrete}.} The length parameters are set to $l_i/M=1$ and $l_f/M=1$.}
\end{figure}

\begin{figure}[!h]
	\centering
	\includegraphics[width=.45\textwidth]{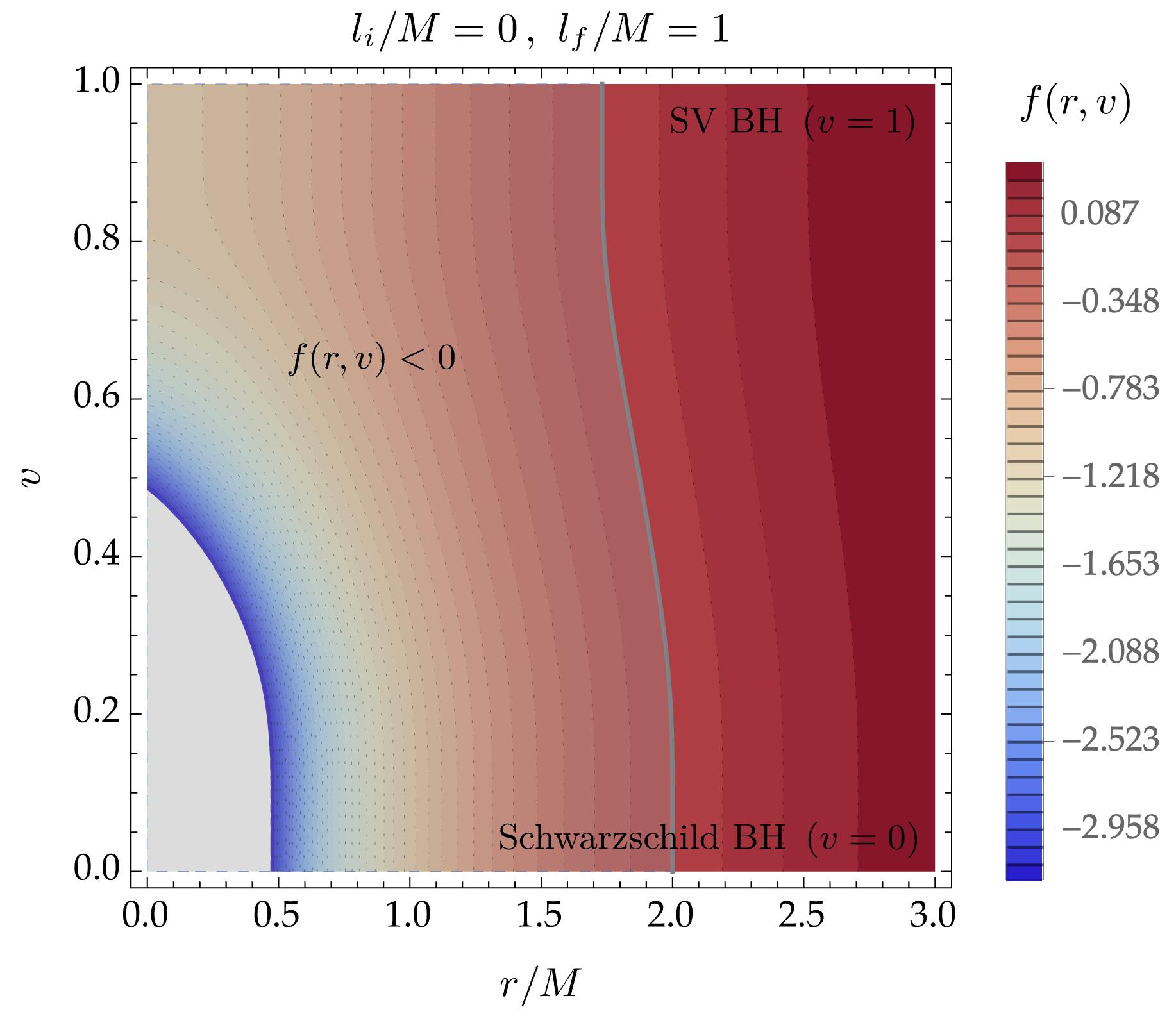}
	\hspace{0.1cm}
	\includegraphics[width=.45\textwidth]{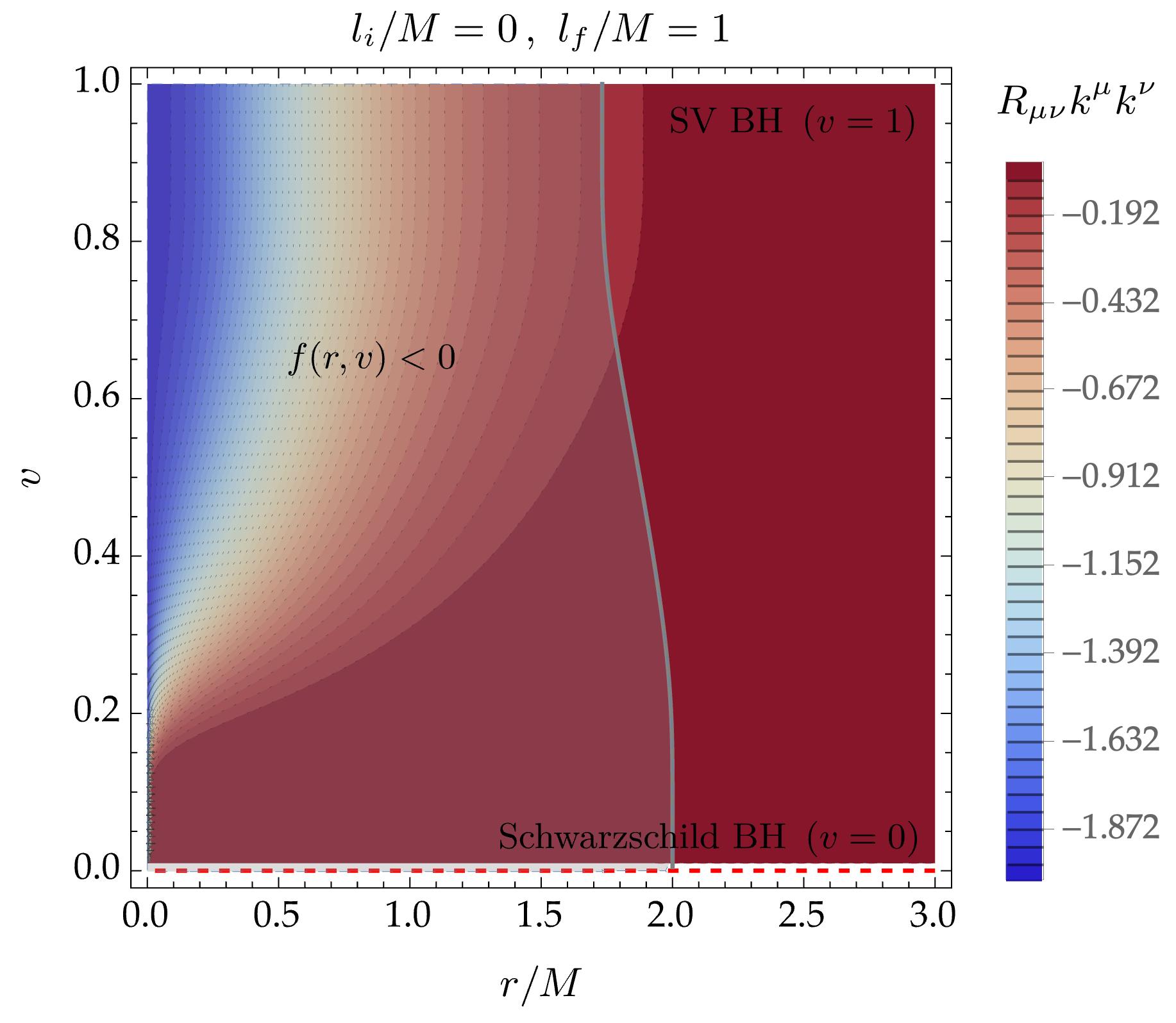}
	\caption{\label{Fig:BHSchwarzschildToSV} Plots of the metric function $f(r,v)$ (left) and the NCC (right) in~\eqref{eq:fNCC2} for a transition from a Schwarzschild black hole to a Simpson--Visser black hole with mass and radial functions~\eqref{eq:MassFunctionsBHSingularSV} and~\eqref{eq:RadialFunctionsBHSingularSV}, via the interpolating mass and radial functions~\eqref{eq:MassFunctionInterpolation} and~\eqref{eq:RFunctionInterpolation}. The gray-shaded region marks the trapped region where $f(r,v)<0$. The uncolored region in the left plot is bounded by the minimum value of the legend and represents a region where the lapse function $f(r,v)$ becomes large negative and off-scale.{Along the red dashed horizontal line $R_{\mu\nu}k^\mu k^\nu = 0$, and the NCC is satisfied. Everywhere else the NCC is violated, in accordance with equation~\eqref{eq:NCCRconcrete}.} The length parameters are set to $l_i/M=0$ and $l_f/M=1$.}
\end{figure}

Fig.~\ref{Fig:BHIntegrableToSV} shows the expressions~\eqref{eq:fNCC2} for an initial geometry with parameter $l_i$ such that the spacetime contains a black hole with an integrable singularity and a final parameter $l_f$ such that the spacetime contains a Simpson--Visser black bounce. Fig.~\ref{Fig:BHSchwarzschildToSV} shows the same quantities for $l_i \to 0$ when the initial geometry is Schwarzschild.

\enlargethispage{100pt}

\clearpage
\subsubsection{BH with integrable or Schwarzschild singularity $\longrightarrow$ SV naked wormhole}

\begin{figure}[!h]
	\centering
	\includegraphics[width=.45\textwidth]{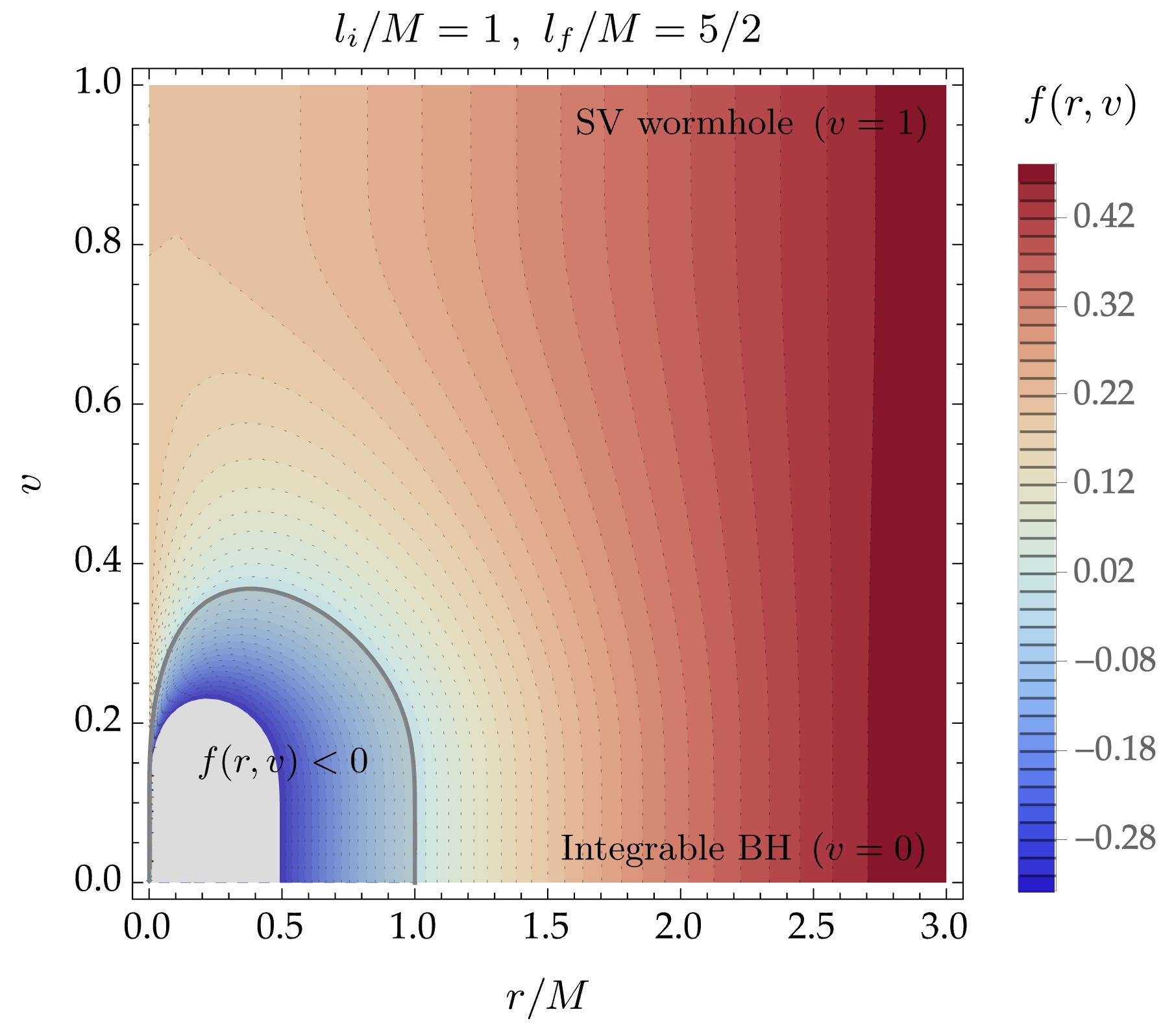}
	\hspace{0.1cm}
	\includegraphics[width=.45\textwidth]{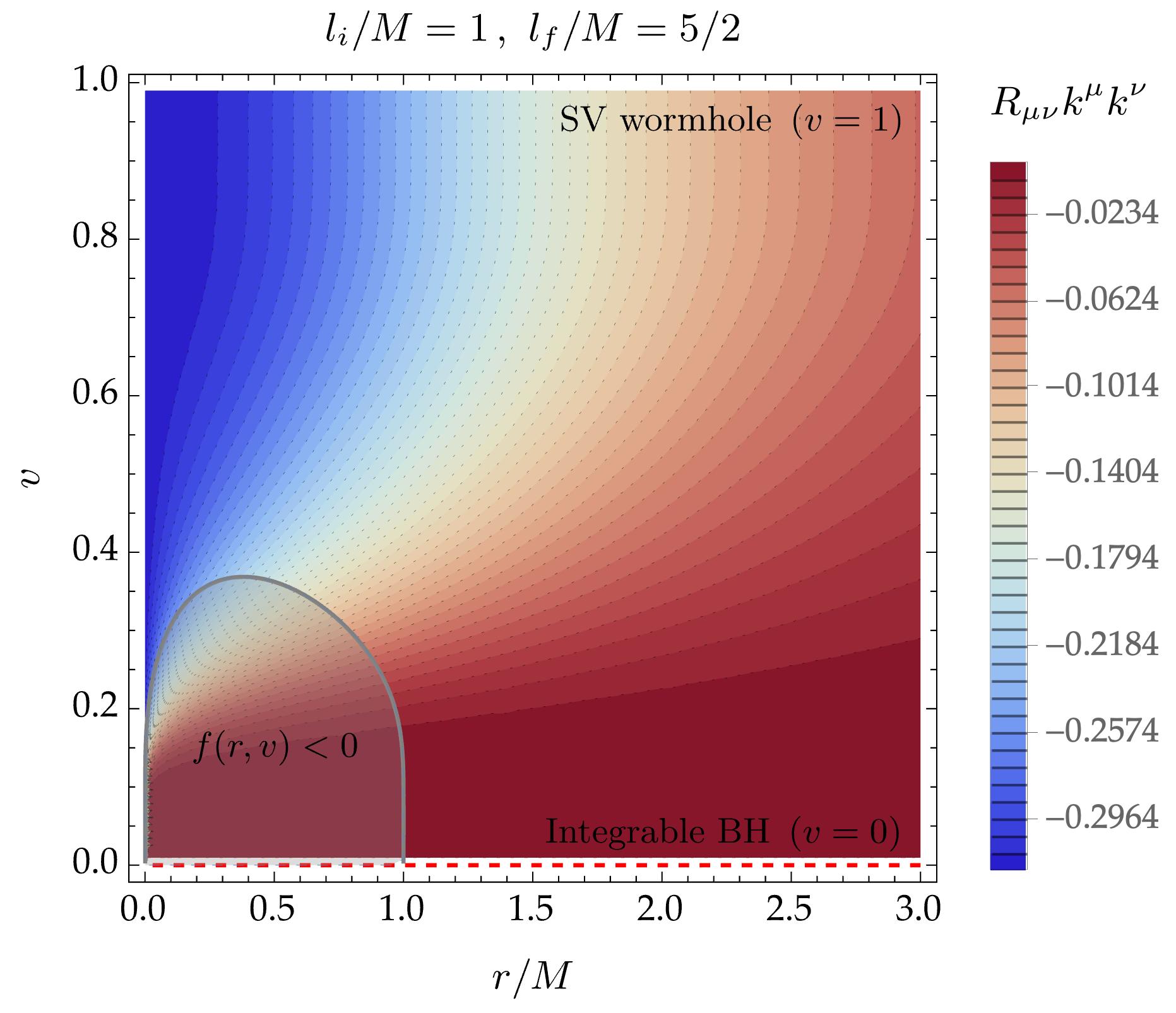}
	\caption{\label{Fig:BHIntegrableToSV2} Plots of the metric function $f(r,v)$ (left) and the NCC (right) in~\eqref{eq:fNCC2} for a transition from a black hole with integrable singularity to a Simpson--Visser naked wormhole with mass and radial functions~\eqref{eq:MassFunctionsBHSingularSV} and~\eqref{eq:RadialFunctionsBHSingularSV}, via the interpolating mass and radial functions~\eqref{eq:MassFunctionInterpolation} and~\eqref{eq:RFunctionInterpolation}. The uncolored region in the left plot is bounded by the minimum value of the legend and represents a region where the lapse function $f(r,v)$ becomes large negative and off-scale. The gray-shaded region marks the trapped region where $f(r,v)<0$. {Along the red dashed horizontal line $R_{\mu\nu}k^\mu k^\nu = 0$, and the NCC is satisfied. Everywhere else the NCC is violated, in accordance with equation~\eqref{eq:NCCRconcrete}.} The length parameters are set to $l_i/M=1$ and $l_f/M=5/2$.}
\end{figure}

\begin{figure}[!h]
	\centering
	\includegraphics[width=.45\textwidth]{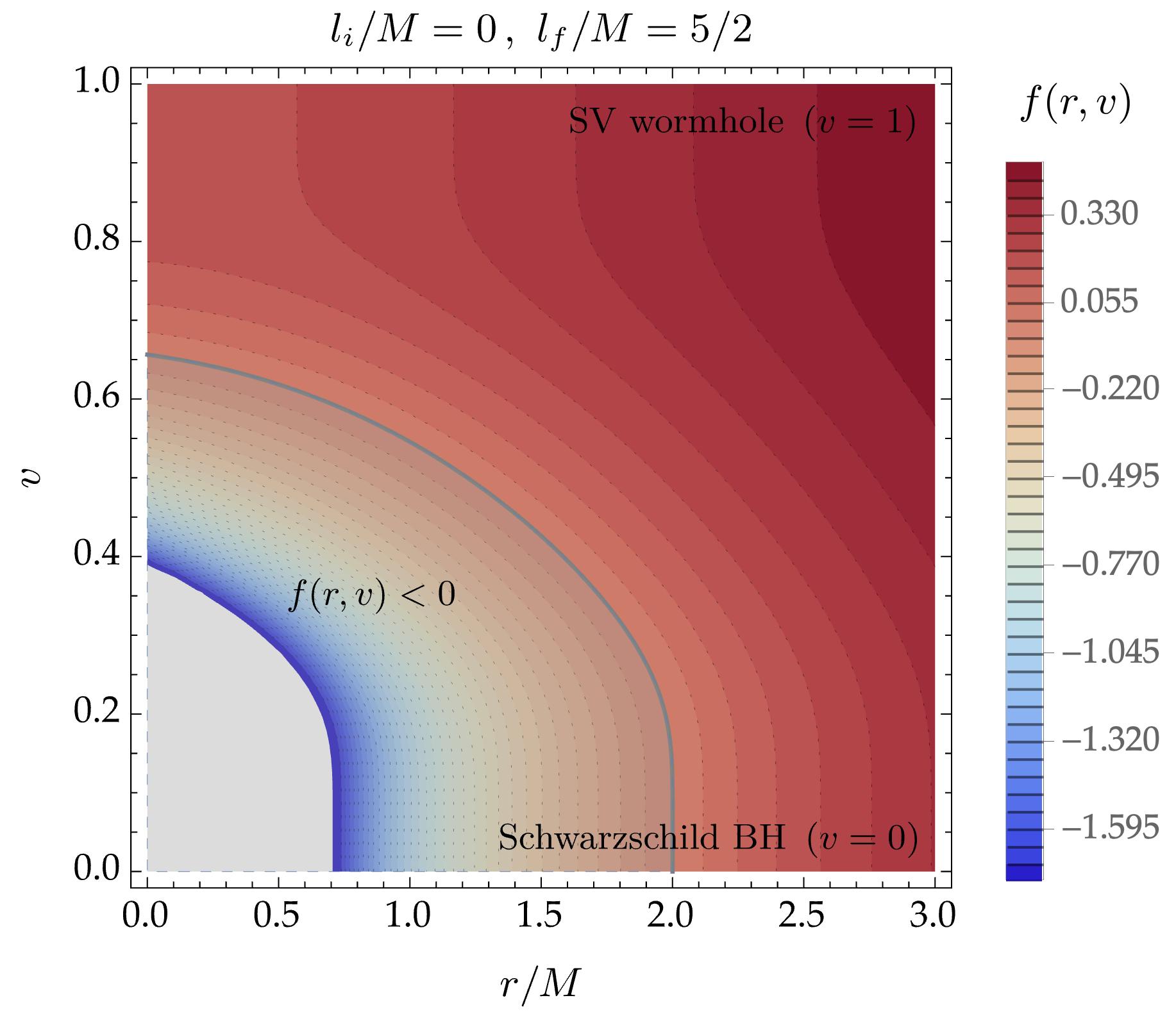}
	\hspace{0.1cm}
	\includegraphics[width=.45\textwidth]{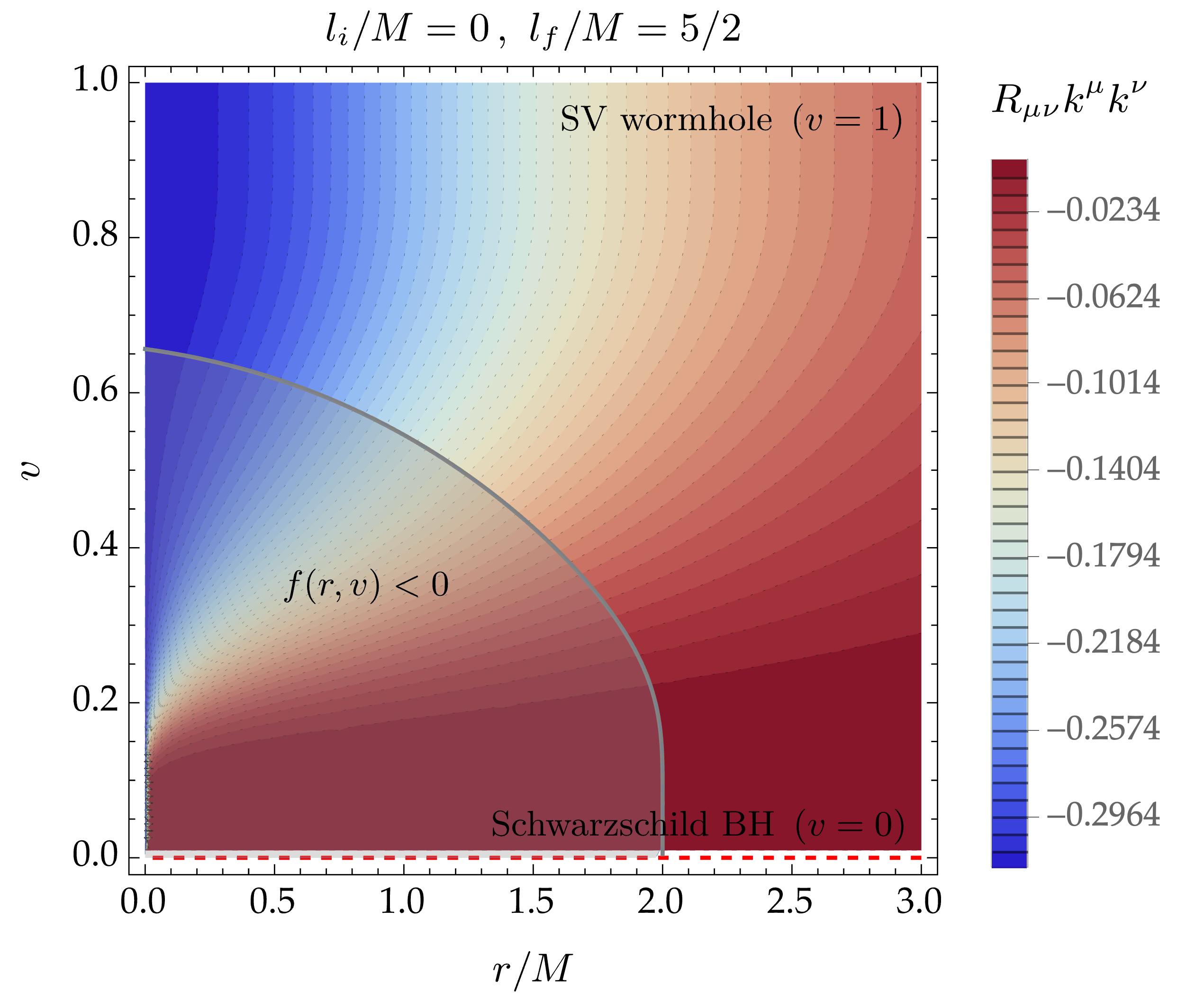}
	\caption{\label{Fig:BHSchwarzschildToSV2} Plots of the metric function $f(r,v)$ (left) and the NCC (right) in~\eqref{eq:fNCC2} for a transition from a Schwarzschild black hole to a Simpson--Visser black hole with mass and radial functions~\eqref{eq:MassFunctionsBHSingularSV} and~\eqref{eq:RadialFunctionsBHSingularSV}, via the interpolating mass and radial functions~\eqref{eq:MassFunctionInterpolation} and~\eqref{eq:RFunctionInterpolation}. The gray-shaded region marks the trapped region where $f(r,v)<0$. The uncolored region in the left plot is bounded by the minimum value of the legend and represents a region where the lapse function $f(r,v)$ becomes large negative and off-scale. {Along the red dashed horizontal line $R_{\mu\nu}k^\mu k^\nu = 0$, and the NCC is satisfied. Everywhere else the NCC is violated, in accordance with equation~\eqref{eq:NCCRconcrete}.} The length parameters are set to $l_i/M=0$ and $l_f/M=5/2$.}
\end{figure}

Fig.~\ref{Fig:BHIntegrableToSV2} shows the expressions~\eqref{eq:fNCC2} for an initial geometry with parameter $l_i$ such that the spacetime contains a black hole with an integrable singularity and a final parameter $l_f$ such that the spacetime contains a Simpson--Visser naked wormhole. Fig.~\ref{Fig:BHSchwarzschildToSV2} shows the same quantities for $l_i \to 0$ when the initial geometry is Schwarzschild.

\enlargethispage{75pt}

\clearpage
\subsection{Transitions from bouncing geometries to wormhole geometries}\label{Sec:SVBHtoSVWormhole}

In the following, we provide examples for the violation of the NCC according to equation~\eqref{eq:NCCk} from an initial geometry describing a regular bouncing geometry with length parameter $l_i$ to a final geometry of the same type but with different values of this length parameter such that the spacetime describes a regular naked wormhole, whose length parameter we denote by $l_f$. To that end we use the interpolating mass function {defined as in equation~\eqref{eq:MassFunctionInterpolation}} together with the interpolating radial function~\eqref{eq:RFunctionInterpolation2}.
For concreteness, we focus on a transition model based on a Simpson--Visser geometry \cite{Simpson:2018tsi,Lobo:2020ffi,Franzin:2021vnj,Lobo:2020kxn}.

\subsubsection{SV black bounce $\longrightarrow$  SV naked wormhole}

\begin{figure}[!h]
	\centering
	\includegraphics[width=.45\textwidth]{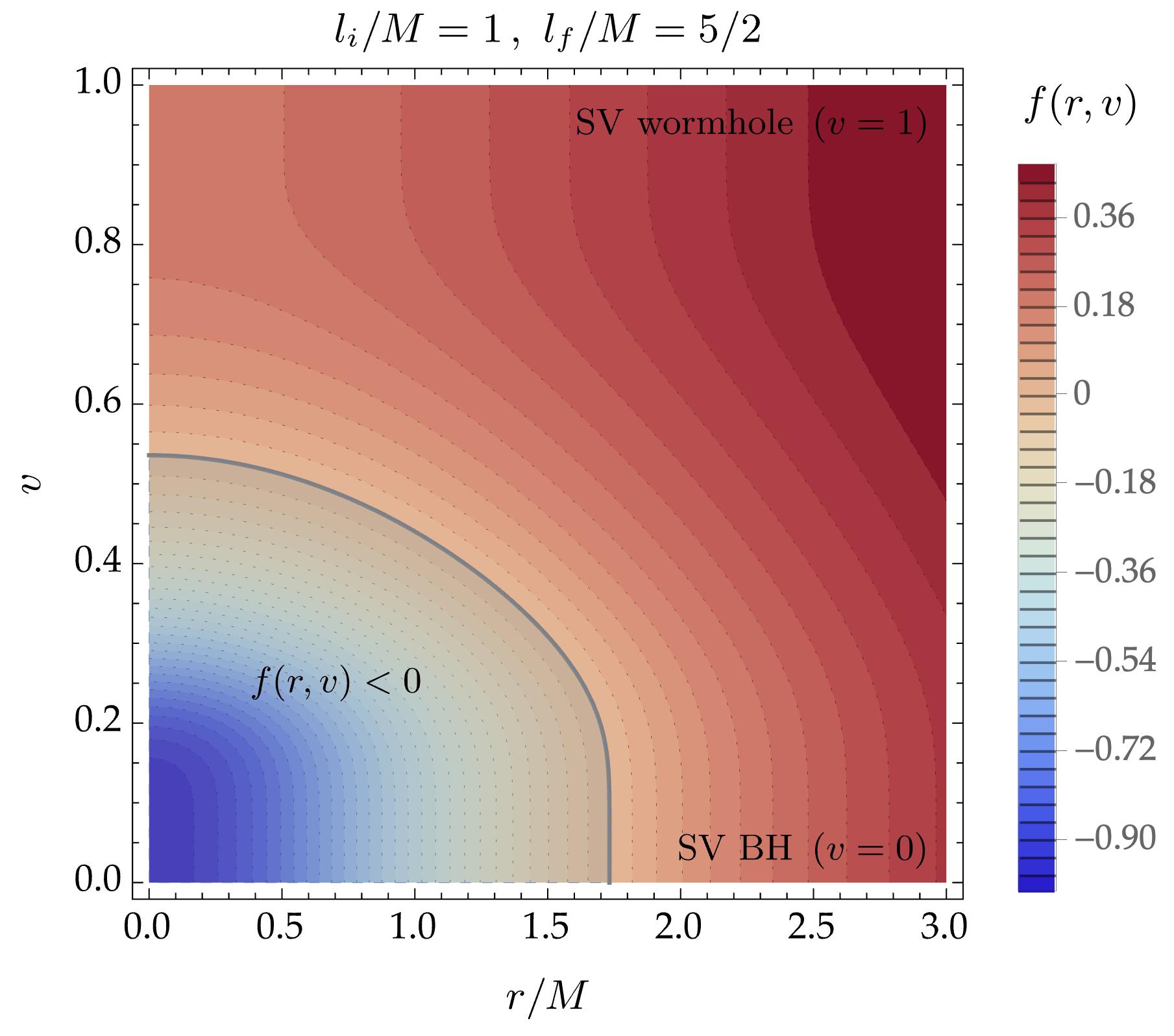}
	\hspace{0.1cm}
	\includegraphics[width=.45\textwidth]{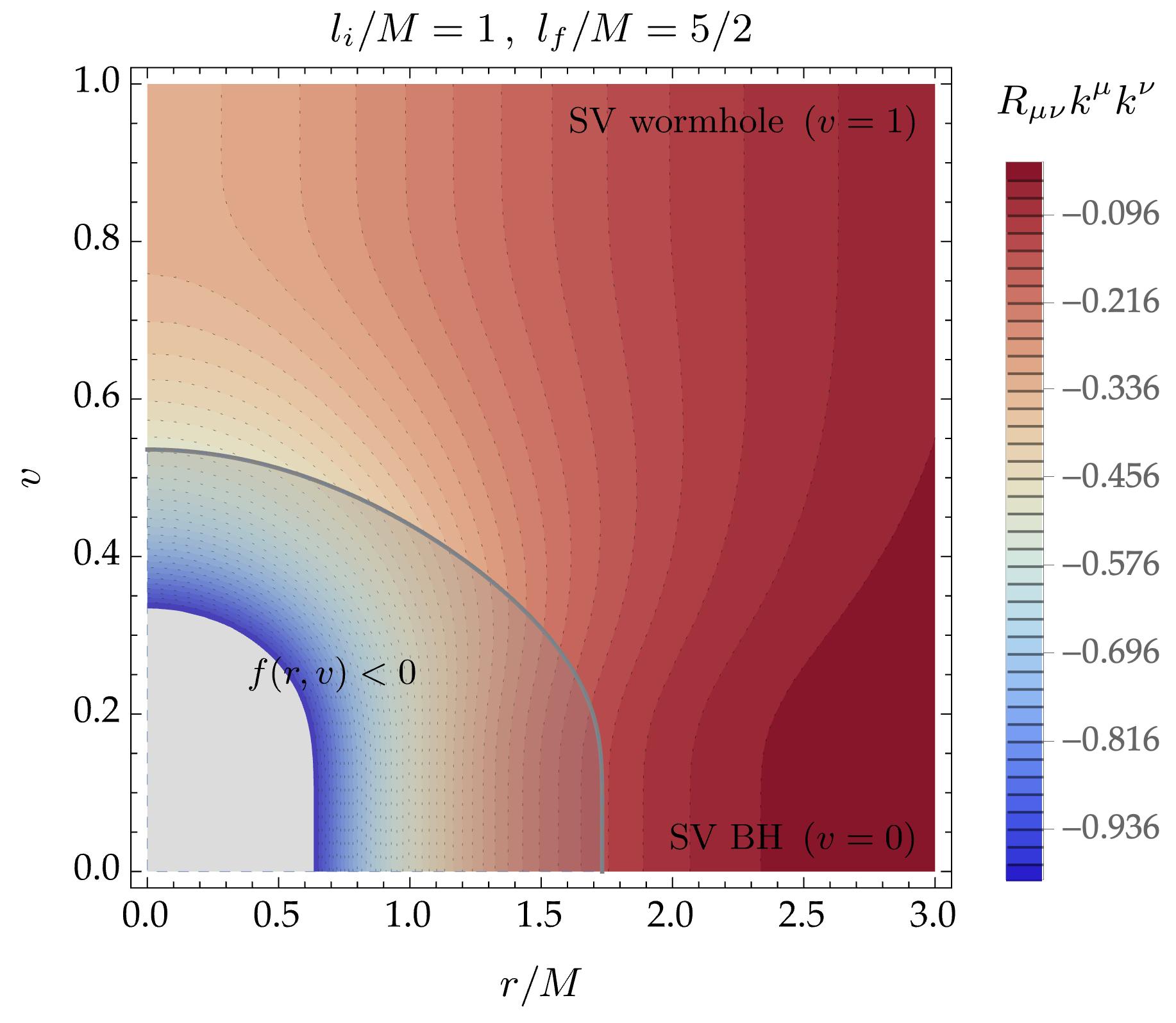}
	\caption{\label{Fig:SVBounceToSVWormhole} Plots of the metric function $f(r,v)$ (left) and the NCC (right) in~\eqref{eq:fNCC2} for a transition from a Simpson--Visser black hole to a Simpson--Visser naked wormhole with mass functions~\eqref{eq:MassFunctionsSVBHToSVWormhole} via the interpolating mass and radial functions~\eqref{eq:MassFunctionInterpolation} and~\eqref{eq:RadialFunctionSVBHToSVWormhole}. The gray-shaded region marks the trapped region where $f(r,v)<0$. The uncolored region in the right plot is bounded by the minimum value of the legend and represents a region of extremely high and off-scale NCC violation. The length parameters are set to $l_i/M=1$ and $l_f/M=5/2$.}
\end{figure}

For the transition from a Simpson--Visser black-bounce spacetime to a Simpson--Visser wormhole spacetime~\cite{Simpson:2018tsi,Lobo:2020ffi,Franzin:2021vnj,Lobo:2020kxn} we use the interpolating mass function $m(r,v)$ in~\eqref{eq:MassFunctionInterpolation} and set
\be\label{eq:MassFunctionsSVBHToSVWormhole}
m_{i}(r) = m_f(r) = M\,.
\ee
{Moreover, according to equation~\eqref{eq:RFunctionInterpolation2}} we set the radial function $R(r,v)$ to be
\be\label{eq:RadialFunctionSVBHToSVWormhole}
R(r,v) = \sqrt{r^2 + l^2(v)}\,,
\ee
where $l^2(v)$ is defined in~\eqref{eq:LengthFunctionInterpolation}.
As before, the quantities of interest are the metric function $f(r,v)$ defined in~\eqref{eq:MetricImplodingSphericalSymmetry} and the left hand side of the NCC in equation~\eqref{eq:NCCk} as summarized in equation~\eqref{eq:fNCC2}. Fig.~\ref{Fig:SVBounceToSVWormhole} shows these expressions for an initial geometry with parameter $l_i$ such that the spacetime contains a Simpson--Visser black hole and a final parameter $l_f$ such that the spacetime contains a Simpson--Visser naked wormhole.
Note that now both the initial and final spacetimes violate the NCC.
So it is now not too surprising that the interpolating spacetime also violates the NCC.

\clearpage

\section{Discussion}\label{Sec:Discussion}

So how do regular black holes evade the Hawking--Penrose~\cite{Penrose:1964wq, Hawking:1970zqf} singularity theorems (see also~\cite{Senovilla:2014gza,Hawking:1973uf})? In this work we have addressed different aspects of this question focussing on the Penrose (1965) theorem~\cite{Penrose:1964wq}, whereas the Hawking--Penrose (1970) theorem~\cite{Hawking:1970zqf} will be discussed elsewhere~\cite{ToAppear}.

As pointed out, for various stationary regular black hole models, it is the inner horizon, which is also a Cauchy horizon, that violates one of the assumptions of the Penrose singularity theorem, despite these models satisfying the NCC globally.

{So, how do regular black holes {actually} form? One possibility is simply that the Penrose singularity theorem cannot be applied to {static or stationary regular black-hole gemoetries}, as they are endowed, (in the same way as the singular Reissner--Nordstr\"om spacetime), with inner horizons which are also Cauchy horizons, and as such violate global hyperbolicity.}
This would, however, be a quite debatable resolution of the conundrum, as such geometries have to be considered as more mathematical abstractions, asymptotic states, rather than realistic models of the end points of a gravitational collapse. Instead, it is reasonable to expect that the Hawking--Penrose (1970) singularity theorem provides insights into this question. The latter assumes the TCC, in addition to the NCC, but relaxes the assumption of existence of a non-compact Cauchy hypersurface. In particular, this theorem establishes the geodesic incompleteness of the Reissner--Nordstr\"om spacetime and exemplifies how de Sitter core regular black holes avoid a singularity by violating the TCC.

Regarding the Penrose singularity theorem, on the other hand, in an evolving time-dependent situation, an inner horizon does not necessarily need to be a Cauchy horizon, and thereby could be compatible with the assumption of global hyperbolicity. In such a situation, it seems that one would be forced to violate the NCC in order to reach a singularity-free endpoint. 

On these grounds we have developed several kinematic models that interpolate between ``standard physics'' (standard black holes or horizonless objects) and regular black holes (assuming spherical symmetry).  and have explicitly verified that during the purely kinematic evolution process (with no particular dynamics is assumed) violations of the NCC are common and seemingly unavoidable, although a formal proof of the universality of this behaviour would certainly be desirable. 

Let us emphasize the broad implications and placement of our classical analysis within the context of the formation of regular black holes and bouncing spacetimes in dynamical situations, and more generally within the context of the quantum theory of such regular objects. The NCC violation required for the transition from a singular initial state to a regular final state in a physically realistic scenario can expected to be induced by quantum effects. A complete description and understanding of this NCC violation would require knowledge of the renormalized stress-energy tensor in dynamical situations.
	
One may consider the viewpoint that NCC violation is secondary to the regularization of black holes, and that it is instead the violation of the Generalized Second Law (GSL) which plays a central role for the quantum theory of regular black holes~\cite{Wall:2010jtc}. Such quantum singularity theorems are however equally inapplicable as the classical singularity theorems for describing the transition to a regular stationary regular black-hole geometry, for which the assumption of global hyperbolicity is violated. One may further object that in physically realistic dynamical situations no Cauchy horizon is present which would violate global hyperbolicity. However, in such situations the applicability of quantum singularity theorems which make use of the GSL is not obvious, due to lack of suitable notions of entropy and theremodynamics quantities in an out-of-equilibriium context.

Let us return to the question of the Cauchy horizon in dynamical situations, and under the impact of quantum effects. It is well-known that Cauchy horizons are subject to a classical instability~\cite{Carballo-Rubio:2018pmi,Carballo-Rubio:2021bpr,DiFilippo:2022qkl}, which notably extends to a classical  instability of dynamical inner horizons~\cite{Carballo-Rubio:2024dca}.

In addition to the classical instability of Cauchy horizons, there is also a quantum instability arising from the incompatibility between the Unruh state and the presence of stationary inner horizons~\cite{Hollands:2019whz}, see also~\cite{McMaken:2024tpc,McMaken:2024fvq,Balbinot:2023vcm}. This illustrates that once a regular black hole has formed, its inner horizon will be subject to an evolving instability. At the present stage the outcome of this evolution is still unknown and moreover it is unclear how such a semiclassical instability could be extended to dynamical inner horizons. Knowledge of the endpoint of such instabilities would in principle allow us to apply our method to this particular final state and determine the renormalized stress-energy tensor (RSET) required to evolve from a singular solution to this endpoint configuration. In the absence of this information we have limited our investigations to the formation of a regular black hole, demonstrating that the interpolation between singular and non-singular solutions necessarily entails a transient violation of the NCC.

Finally let us come back to the classical aspects of our analysis. We feel that this investigation serves to highlight more than ever the pressing need for improved singularity theorems better-adapted to dealing with evolving inner and outer trapping horizons~\cite{Hawking:2014tga,Visser:2014zqa,Ashtekar:2004cn}, as well as an improved understanding of existent singularity theorems and their global and geometric convergence conditions. In particular, our investigations emphasize the need {for} an improved understanding of general conditions under which the formation of timelike singularities --- shielded by dynamical inner horizons --- is unavoidable.
{In conclusion, one might say that also in this situation}, as has recently happened for mass inflation~\cite{Carballo-Rubio:2024dca}, regular black holes are forcing us to {carefully} reconsider the dynamical aspects of trapped regions and their horizons. As such, independently of their phenomenological interest, they ought to be considered crucial objects for current and future research in black hole physics (see e.g.~\cite{Carballo-Rubio:2023mvr,DiFilippo:2024spj, Buoninfante:2024oxl,Visser:2024zkx, Carballo-Rubio:2025fnc}).

\begin{acknowledgments}
	JB is supported by a doctoral scholarship by the German Academic Scholarship Foundation (Studienstiftung). Research at Perimeter Institute is supported in part by the Government of Canada through the Department of Innovation, Science and Economic Development Canada, and by the Province of Ontario through the Ministry of Colleges and Universities.\\
    During early phases of this work MV was supported by a Victoria University travel grant, and by SISSA and INFN.
\end{acknowledgments}

\enlargethispage{20pt}
\bibliographystyle{jhep}
\bibliography{referencesNCC}

\end{document}